# Carbon-based Microfabricated Organic Electrochemical Transistors Enabled by Printing and Laser Ablation

*Alan Eduardo Avila Ramirez,§ Jessika Jessika,§ Yujie Fu, Gabriel Gyllensting, Marine Batista, David Hijman, Jyoti Shakya, Yazhou Wang, Wan Yue, Renee Kroon, Jiantong Li, Mahiar Max Hamedi, Anna Herland,\* Erica Zeglio\**

A.E. Avila Ramirez, J. Jessika, A. Herland
Division of Nanobiotechnology, SciLifelab, Department of Protein Science, KTH Royal Institute of Technology, Tomtebodavägen, 23a, 171 65, Solna, Sweden

Y. Fu, J. Li
School of Electrical Engineering and Computer Science, KTH Royal Institute of Technology, Electrum 229, Kista 16440, Sweden

G. Gyllensting, M. Batista, D. Hijman, E. Zeglio
Wallenberg Initiative Materials Science for Sustainability, Department of Chemistry, Stockholm University, 114 18, Stockholm, Sweden

M. Batista, A. Herland, E. Zeglio
AIMES – Center for the Advancement of Integrated Medical and Engineering Sciences at Karolinska Institutet and KTH Royal Institute of Technology, Stockholm, Sweden

J. Shakya, M. M. Hamedi
Department of Fibre and Polymer Technology, KTH Royal Institute of Technology, Teknikringen 56, 10044 Stockholm, Sweden.

Y. Wang
Organic Bioelectronics Laboratory, Biological and Environmental Science and Engineering (BESE) Division, King Abdullah University of Science and Technology (KAUST), Thuwal, Saudi Arabia

W. Yue
State Key Laboratory of Optoelectronic Materials and Technologies, Key Laboratory for Polymeric Composite and Functional Materials of Ministry of Education, Guangzhou Key Laboratory of Flexible Electronic Materials and Wearable Devices, School of Materials Science and Engineering, Sun Yat-sen University, Guangzhou, 510275 P. R. China

R. Kroon

Wallenberg Initiative Materials Science for Sustainability, Laboratory of Organic Electronics, Department of Science and Technology, Linköping University, Norrköping, Sweden.

*Erica Zeglio: erica.zeglio@su.se

*Anna Herland: aherland@kth.se

**Abstract:** Organic electrochemical transistors (OECTs) are key bioelectronic devices, with applications in neuromorphics, sensing, and flexible electronics. However, their microfabrication typically relies on precious metal contacts manufactured via cleanroom processes. Here, we present a high-throughput additive-subtractive microfabrication strategy for metal-free, flexible OECTs using biodegradable materials and room-temperature processing. Additive manufacturing of large features is achieved via extrusion printing of a water-dispersed graphene ink to fabricate electrode contacts, and spin-coating of a cellulose acetate ink to form both the substrate and encapsulation layer. Combined with femtosecond laser ablation, this approach enables micrometer-resolution patterning of free-standing OECTs with channel openings down to 1 μm and sheet resistance below 10 Ω/sq. By tuning laser parameters, we demonstrate both selective and simultaneous ablation strategies, enabling the fabrication horizontal, vertical, and planar-gated OECTs, as well as complementary NOT gate inverters. Thermal degradation studies in air show that over 80% of the device mass decomposes below 360 °C, providing a low-energy route for device disposal and addressing the environmental impact of electronic waste. This approach offers a cleanroom-free and lithography-free pathway toward the rapid prototyping of high-resolution, sustainable organic electronics, combining material circularity, process simplicity, and architectural versatility for next-generation bioelectronic applications.

# 1. Introduction

Organic electrochemical transistors (OECTs) leverage the unique properties of organic mixed ionic-electronic conductors (OMIECs) to combine low-voltage operation (below 1 V), high signal amplification (i.e., transconductance), and straightforward integration into mechanically flexible and conformable systems.[1] These features make OECTs highly suitable for applications in bioelectronics, such as biosensors,[2] neuromorphic circuits,[3] wearable electronics,[4] implantable therapeutics,[5] food packaging,[6] and plant science applications.[7] Despite advances in the molecular engineering of OMIECs and device geometry prototyping, most high-performance OECTs are still fabricated using lithography,[8] which remains the gold standard technique due to its high resolution (micro/nano-scale) and batch-to-batch reproducibility. However, challenges such as high costs and complex multistep processes pose barriers to fast prototyping and further scale up, along with additional fabrication constraints due to incompatibility with biomaterials and flexible and/or degradable substrates.[8]

Additive manufacturing techniques, such as inkjet and screen printing, have emerged as low-cost, maskless alternatives suitable for flexible electronics.[9] Inkjet printing can achieve resolutions in the 20–50 μm range for 2D patterns but is limited by drop placement accuracy, nozzle clogging, material viscosity (4 – 14 cP), density (~1 g cm$^{-3}$), surface tension (29-50 N m$^{-1}$) constraints, and expensive cartridges.[10,11] Screen printing, though highly scalable and industry-ready, offers relatively coarse resolution (40 μm)[12] and requires large ink volumes (from milliliters to deciliters and even liter scale), making it unsuitable for prototyping using novel materials. Despite these limitations, both methods have been widely explored for OECT fabrication, with some approaches combining the two in hybrid workflows for fully printed devices.[13]

Extrusion-based printing emerged as a promising alternative, offering a versatile, maskless, and user-friendly approach for fabricating with a wide range of materials through simple syringe-assisted deposition of various inks. Depending on the needle's inner diameter (i.e., 34 Gauge), the resolution of extrusion printing ranges from 50 μm to 100 μm, while micro extrusion printers can achieve resolutions as fine as 5 μm.[14] However, rheological properties, such as low viscosity and a lack of yield stress, and speed of movement can increase droplet or filament size beyond 200 μm.[15] The technique has been successfully applied in OECT fabrication, including printing silver and carbon electrode contacts,[16] and even for creating fully 3D-printed OECTs using composites like reduced graphene oxide and carbon nanotubes.[17] Yet, additive manufacturing approaches alone still lack the resolution needed to downscale to micrometer resolution (1–10 μm), necessary for defining short channel lengths - critical to achieve higher transconductance, higher integration density, and overall highly-performing devices.[18]

To address this challenge, we have investigated the integration of subtractive manufacturing to complement the rapid additive pattern creation provided by additive methods with precise ablation of critical features. Femtosecond laser writing offers micrometer-scale resolution with minimal thermal damage, making it ideal for processing organic and carbon-based materials. This technique has been employed to pattern silver contacts for organic synaptic transistors.[19] Moreover, we recently demonstrated that direct femtosecond laser writing can be used to pattern insulation and active layers enabling direct, maskless patterning of OECT components with high spatial precision.[20] However, these works used lithography and vacuum evaporation through a mask to fabricate electrode contacts and insulation layers (i.e., Parylene C).

In this work, we integrate additive and subtractive methods to overcome the limitations of each technique alone, with a focus on sustainable materials and room-temperature processes. Our approach combines extrusion printing, spin-coating, and femtosecond laser ablation to produce flexible, carbon-based OECTs. Water-dispersible, 2D nanomaterials, such as graphene ink in water, offer low sheet resistance and long-term stability, making them ideal as electrode contacts for OECTs.[21,22] At the same time, cellulose-based materials, derived from renewable sources, are abundant, biodegradable, and compatible with solvent processing, making them suitable as the substrate and encapsulation layer.[23,24]

Our fabrication protocol enables printed graphene electrodes with sheet resistance below 10 $\Omega$ sq$^{-1}$ and critical dimensions down to 1 μm via femtosecond laser ablation. Entire device stacks remain below 10 μm in thickness, with reliable encapsulation and full delamination to produce flexible, freestanding, and thermally degradable devices. We demonstrate fabrication of horizontal and vertical OECTs, as well as in-plane gated OECTs using printed graphene as the gate. In contrast to conventional microelectrodes for OECTs, made of precious metals (i.e., gold or silver), our devices, made from organic or carbon-based materials, can degrade at lower temperatures, making them a promising alternative to improve the sustainability of organic electronics; therefore, offering a pathway for reducing e-waste.

## 2. Results and Discussion

### 2.1. Cleanroom-free fabrication of encapsulated carbon-based devices

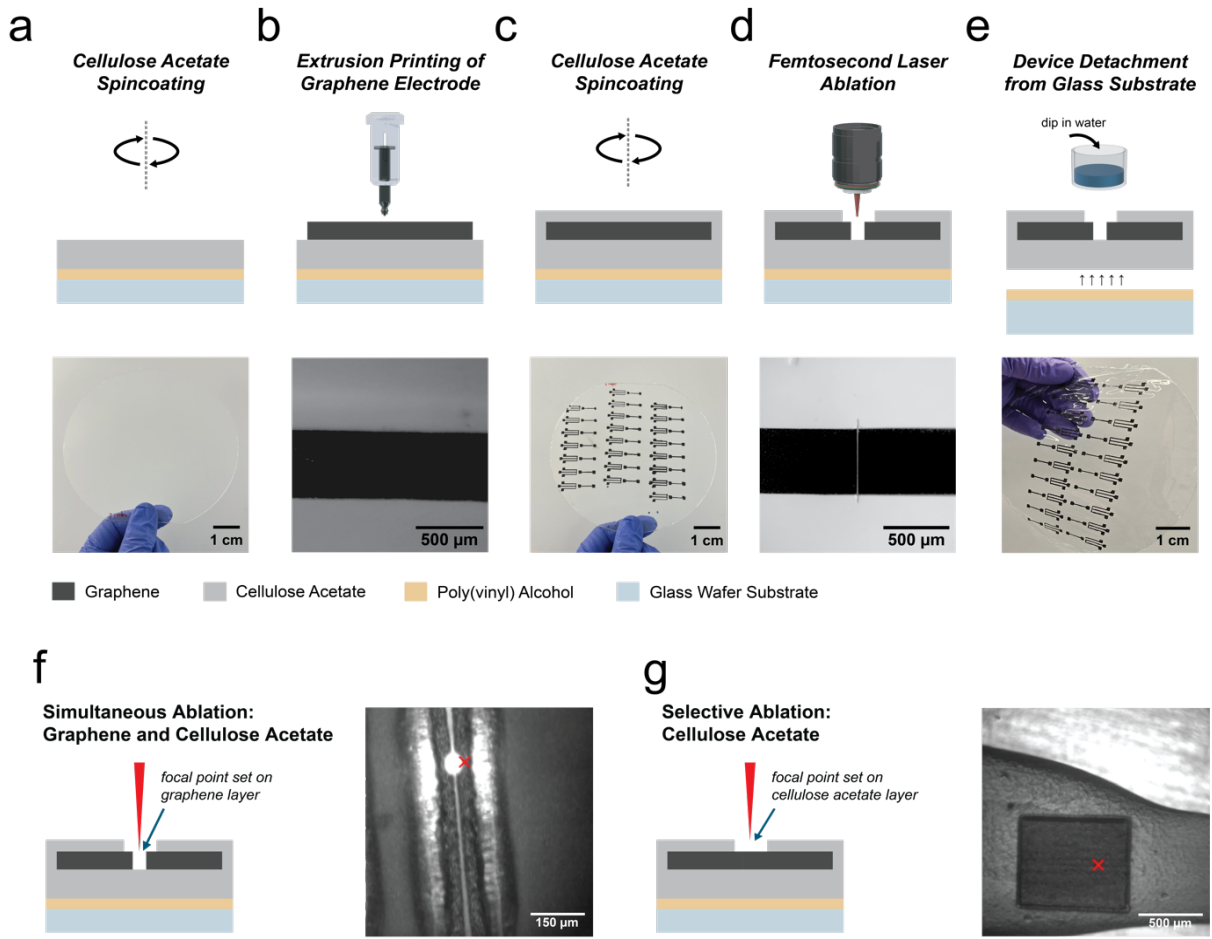

**Figure 1. Additive-subtractive manufacturing of sustainable encapsulated electrodes for OECTs.** Schematic workflow of the fabrication process: (a) A sacrificial polyvinyl alcohol (PVA) layer is first spin-coated onto a glass wafer, followed by spin-coating of the cellulose acetate substrate. (b) Graphene ink is extrusion-printed onto the cellulose acetate surface to define electrode patterns. (c) After drying and annealing, a second cellulose acetate layer is deposited as encapsulation. (d) Femtosecond laser ablation is used to open the transistor channel or contact pad regions. (e) Devices are finally delaminated by dissolving the PVA layer in water, yielding flexible, self-standing electrode arrays. (f-g) Schematic of femtosecond laser ablation strategies. (f) Left: simultaneous ablation of graphene and encapsulating cellulose acetate layer using a 10× objective. Right: preview on the ablation process. (g) Left: selective ablation of only the cellulose acetate encapsulation using a 4× objective. Right: preview on the ablation result. The X in red indicates the cursor inside the screen for the femtosecond-laser camera manipulation.

Our proposed method enables the microfabrication of fully organic, carbon-based OECTs by integrating both additive and subtractive manufacturing techniques. The device stack is constructed using spin-coating and extrusion printing of acetone- and water-based inks, with cellulose acetate serving as both the substrate and insulating layer, and graphene as the electrode material (**Figure 1a-e**). To achieve microscale patterning of the channel area, we employed direct femtosecond laser ablation of the encapsulated graphene and cellulose acetate layers.

We chose the bio-based material cellulose acetate as both the substrate and passivation layer due to its transparency, film-forming properties, mechanical stability, water resistance, and degradability through chemical or enzymatic hydrolysis.[25,26] To facilitate handling during fabrication, a 6-inch glass wafer is used as the support. A water-soluble polyvinyl alcohol (PVA) layer is first spin-coated onto the wafer as a sacrificial underlayer, allowing the devices to be easily peeled off after fabrication through simple dissolution of PVA in water (see Experimental Section for details).

The cellulose acetate substrate is prepared by spin-coating from an acetone solution to create a cellulose acetate film, as shown in **Figure 1a**. Film thickness is determined by adjusting the spin-speed, with typical values ranging from 500 to 2500 RPM, yielding to thicknesses ranging from 9.82±0.36 μm to 3.46±0.30 μm, respectively (**Figure S1a**). These values are comparable to typical thicknesses of OECT substrates, such as Parylene C or polyimide,[27] but with the added benefits of avoiding energy-intensive chemical vapor deposition. For device fabrication, we selected a cellulose acetate thickness of ~5 μm (spin-speed of 1500 rpm) as the substrate, providing sufficient mechanical stability following delamination from the glass wafer (**Figure S1b**).

Prior to printing the electrode contacts, the cellulose acetate film is surface treated with a handheld corona surface treater for 5 minutes to enhance graphene adhesion (see Experimental Section). We then used 7 wt% graphene ink in water to 3D print 20 electrode patterns supporting three different architectures: standard horizontal OECTs with an Ag/AgCl gate, planar OECTs with a printed graphene gate, and vertical OECTs. A single pass through a 200 μm diameter nozzle results in electrode lines with a final linewidth of around 500 μm (**Figure 1b**), consistent with the lateral spreading of the printed pattern during drying at room temperature (1 minute) prior to thermal annealing. Previous reports have shown that annealing improves conductivity and adhesion of printed graphene to glass substrates.[28] We investigated annealing temperatures ranging from 100 °C and 150 °C for 10 minutes, which is within the range of the glass transition temperature of cellulose acetate, but below the melting temperature. Four-point probe measurements confirm a stable sheet resistance of $5.9 \pm 1.1$ $\Omega$ sq$^{-1}$ across this temperature range (**Figure S2a**). Raman spectra reveal a red shift in the D and G bands up to 100 °C, similar to what reported in the literature (**Figure S2b** and **S3**), consistent with the multilayer nature of graphene.[29] No further changes in the Raman spectra were observed beyond that temperature, confirming that the annealing step does not negatively impact the structure of printed graphene films. Based on these results, we chose 130°C as the optimal annealing temperature to ensure effective water removal without damaging the underlying cellulose acetate substrate.

Following electrode printing and thermal annealing, we performed a second spin-coating step to deposit the insulation layer, resulting in a uniform, self-standing, wafer-sized multi-electrode structure (**Figure 1c**, see Experimental Section). A subsequent subtractive patterning step using femtosecond laser ablation was then employed to define the channel area with micrometer resolution – surpassing the resolution limitations of the printed features (**Figure 1d**). Successful channel opening was confirmed via optical microscopy in transmission mode (**Figure 2a**). The final delamination step from the glass wafer yielded flexible, self-standing devices with fully encapsulated graphene electrodes (**Figure 1e**).

To demonstrate the flexibility of our fabrication method, we implemented two device architectures: the conventional horizontal OECT (*h*OECT) and the emerging vertical OECT (*v*OECT). While *h*OECTs are widely used and well-established, *v*OECTs offer key advantages, such as reduced channel dimensions with shorter effective channel lengths and the possibility to increase device density via vertical stacking.[30,31] Both architectures were fabricated with diluted PEDOT:PSS as the channel material to realize depletion-mode OECTs. Schematics of the respective fabrication steps are shown in **Figure 1f** and **Figure 1i.**

## 2.2. Simultaneous graphene–cellulose acetate ablation for horizontal OECTs

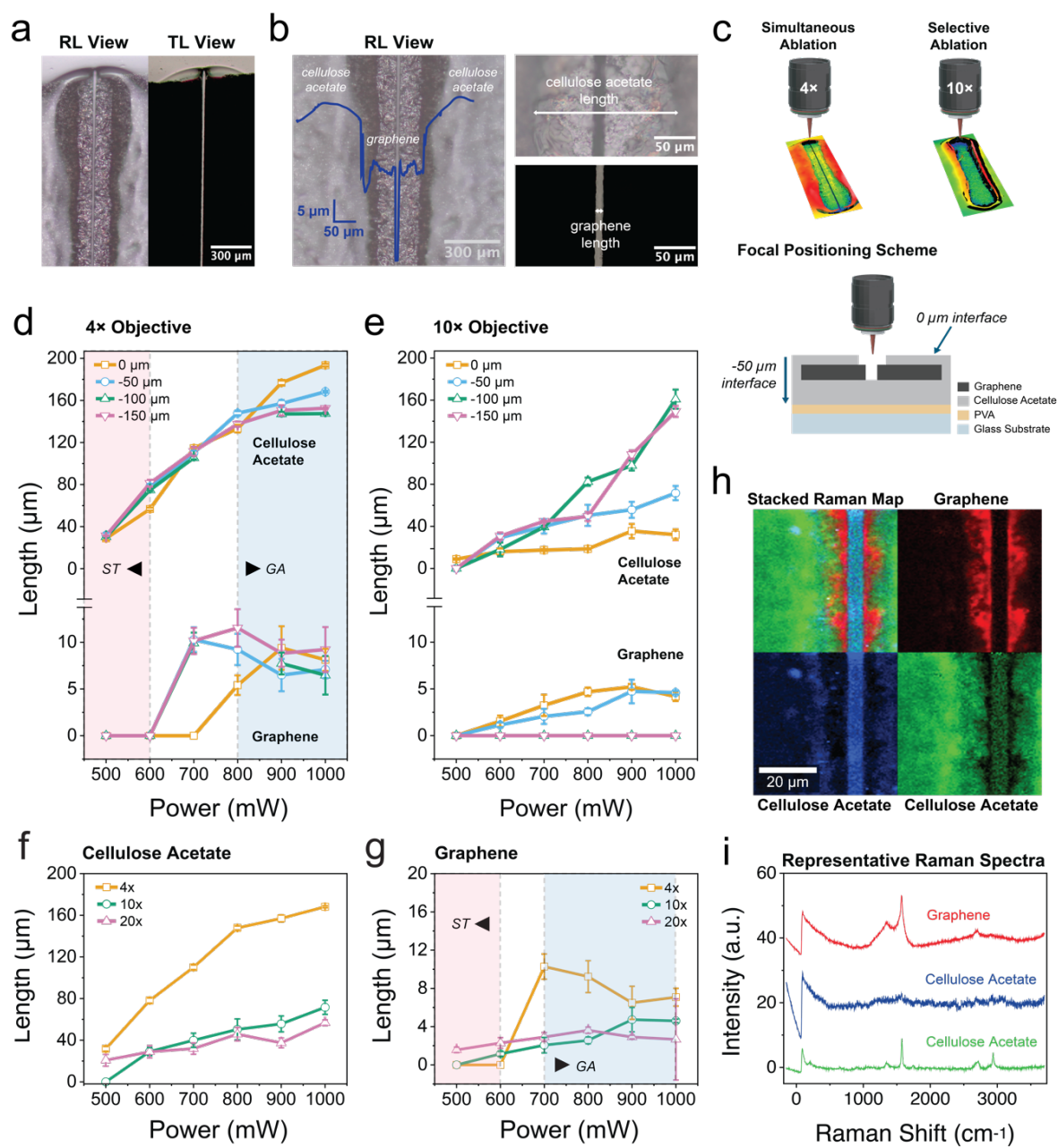

**Figure 2. Femtosecond laser ablation parameters for single-pass simultaneous patterning of graphene and cellulose acetate.** (a) Successful patterning was verified by upright optical microscopy, with corresponding reflecting (RL) and transmitting (TL) light images showing the opened channel. (b) On the left, microscope image (RL) and Z-height profile obtained from 3D optical profilometry (in blue) showing the surface morphology of cellulose acetate and underlying graphene across the ablated region. On the right, magnified reflecting (top right) and transmitting (bottom right) images showing the channel opening following a single pass laser ablation, with the exposed graphene and cellulose acetate lengths quantified in the next plots. (c) On the top, profilometry scans for simultaneous or selective ablation, and schematic

illustration of the focal positioning strategy. Ablation depth is tuned by adjusting the Z-height from 0 to −150 μm. On the bottom, schematic illustration of the femtosecond laser ablation process using a 4× objective for simultaneous removal of cellulose acetate layer and underlying graphene. (d–e) Lateral opening dimensions for (d) cellulose acetate and (e) graphene at varying Z-heights (0 to −150 μm), confirming depth-dependent ablation behavior. The graphene ablation profile marked as GA indicates the region where both cellulose acetate encapsulation and graphene are ablated, while ST is an indication for subthreshold region, where only the cellulose acetate encapsulation layer is ablated, but not graphene. (f–g) Quantified lateral opening lengths in (f) cellulose acetate and (g) graphene as a function of laser power at different objective magnifications. (h) Confocal Raman microscopy of laser-ablated regions following simultaneous ablation with 10× objective, showing spatial separation between graphene (red) and cellulose acetate (green/blue). (i) Representative Raman spectra extracted from confocal Raman data in (h), confirming material identity and quality: graphene (D, G, 2D peaks) and characteristic cellulose acetate signatures from the encapsulating and substrate layers.

We systematically investigated how laser processing parameters influence the lateral opening profiles of both cellulose acetate and graphene layers during a single, simultaneous femtosecond laser ablation step (analyzed via 3D optical profilometry, **Figure S4**). Parameters evaluated include laser pulse power, focal offset (Z-height), objective magnification, and scanning speed (**Figure 2a-g**, **Figure S4-S8**, Supporting Information).

The focal point position (Z-height) plays a critical role in determining which layer is effectively ablated **(Figure 2d-e).** We restricted the focal point variation to a Z-height range down to -150 μm, as further defocusing no longer influences the topmost surface. At 4× magnification, we observed simultaneous ablation of both graphene and cellulose acetate encapsulation layer across a broad Z-range (0 to -150 μm, **Figure 2d**). This is due to the larger depth of field of the 4× objective, distributing the laser fluence across a broader volume. In contrast, at 10× magnification we observe graphene ablation only within a limited focal offset range, down to approximately -50 μm from the cellulose acetate surface, whereas cellulose acetate exhibits a broader ablation tolerance down to -150 μm (**Figure 2e**). Beyond -100 μm, laser energy becomes insufficient to ablate graphene regardless of the power level, indicating the existence of a focal threshold specific to graphene. At 20× magnification, graphene ablation occurs at a very narrow focal range (Z-height at -50 μm) and became highly sensitive to alignment, making it more complicated to use in a reliable manner (data not shown).

Increasing the laser power from 500 to 1000 mW at uniform focal point of -50 μm results in a wider lateral opening in both materials (**Figure 2f,g**). After a single pass of simultaneous laser ablation, cellulose acetate exhibits a consistent expansion in opening length from around 30 μm to over -150 μm, while graphene opening lengths range from around 1 to 12 μm depending on the magnification of the objective. Ablation performed with a 4× objective produces greater variability in graphene opening length, whereas 10× and 20× objectives yield more gradual and controlled increase (**Figure 2f,g**). These results demonstrate that the 10× objective, with its narrower depth of field than the 4× objective, exhibits a more selective ablation of either cellulose acetate encapsulation layer or both cellulose acetate and graphene layers.

Based on these data, we selected the 10× objective for all subsequent simultaneous ablation processes, as it provides an optimal balance between spatial resolution, fabrication robustness, and ease of operation. The lateral opening profile across all three objectives confirms that the 20× objective offers only minimal resolution improvements over the 10× objective, while significantly complicating focus adjustment (**Figure 2g**). The 10× magnification configuration thus offers improved spatial resolution and ablation control compared to 4×, without the alignment challenges associated with 20× magnification.

Scanning speed and number of passes were further investigated to assess the robustness of the simultaneous ablation process. While the lateral opening dimensions remain relatively consistent across different conditions **(Figure S5,** Supporting Information), slower scanning speeds results in higher local laser fluence, leading to pronounced vertical bulging along the sidewalls of the surrounding cellulose acetate **(Figure S8,** Supporting Information). This bulging is attributed to excessive heat accumulation, which causes localized melting and flow-induced material deformation at the edges of the ablated region. As the scanning speed increases, the extent of bulging decreases, indicating that shorter dwell times reduce thermal buildup and allow for more efficient heat dissipation. Thereby, we recommend a scanning speed of 1000-3000 μ ms$^{-1}$ as an optimal trade-off between minimal thermal bulging effect and process throughput.

We observed similar fabrication effects when comparing single-pass and double pass ablation at varying scan speed **(Figure S6-S7,** Supporting Information). The additional energy delivered during the second pass has minimal impact on the lateral opening of cellulose acetate and graphene, as the ablation threshold had already been reached in the initial pass. Since the second pass does not significantly increase the lateral ablation dimensions, it proves beneficial in reducing residual cellulose acetate on the graphene surface. Thus, performing double passes is advisable to help improving surface cleanliness without affecting the lateral opening. Notably, we found that the presence of graphene enhances the efficiency of cellulose acetate ablation. The different ablation responses of graphene and cellulose acetate to single-pass laser exposure are attributed to the distinct ablation mechanisms of the two materials: non-linear optical absorption in graphene versus thermally driven ablation in cellulose acetate (see **Note S1** and **Figure S9** Supporting Information).

To further complement the information from static microscopy and profilometry analyses, we include **Supporting Videos S1-S4** that visualize the femtosecond laser ablation process in real time. These videos demonstrate both simultaneous ablation of graphene and cellulose acetate (**Video S1-S3**) and selective ablation of the cellulose acetate layer alone (**Video S4**). They highlight dynamic factors such as laser pass number, subthreshold optimization, and the visual cues such as channel darkening and illumination angle that assist in verifying successful ablation (see Supporting Information, Video Captions for detailed annotations).

We used confocal Raman microscopy of the patterned regions to validate material selectivity and for channel formation. Raman mapping data confirms the successful formation of the channel, revealing exposed cellulose acetate substrate within the channel region and graphene contacts on both sides. The Raman spectra display the characteristic signatures of graphene (D, G, and 2D peaks) and cellulose acetate, confirming that the fabrication process preserves the integrity of the materials across the different layers, including the insulation layer above the graphene contacts. (**Figure 2h-i**). These results support further use of simultaneous graphene-cellulose acetate ablation for channel formation in horizontal OECTs.

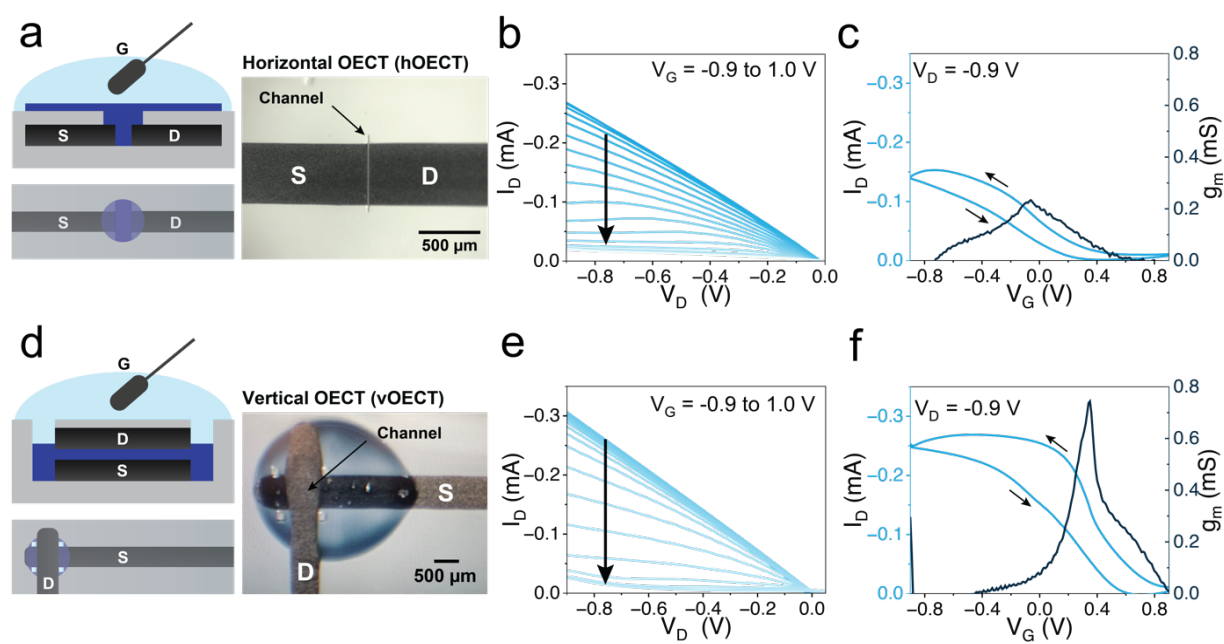

**Figure 3. Fabrication approach and characterization of different OECT architectures.** (a) Device schematic represented by the (top) cross-sectional view and (bottom) top view, with microscope image of an ablated drain-source opening used for horizontal OECT (*h*OECT) architecture. (b) Representative output characteristics of *h*OECT device. (c) Representative transfer characteristics and transconductance of *h*OECT (*W = 500 μm, L = 1 μm*). (d) Vertical OECT (*v*OECT) architecture, with (top) cross-sectional view and (bottom) top view, also the microscope image of a fully fabricated *v*OECT. (e) Representative output characteristics of *v*OECT, with ablation performed after the encapsulation of electrodes and the PEDOT:PSS channel. (f) Representative transfer characteristics and transconductance of *v*OECTs (*W = 500 μm, L = drop-casted PEDOT:PSS, and d ≈ 525 μm*). These measurements were performed using 1X PBS electrolyte and an Ag/AgCl pellet as gate electrode.

Although fully additive manufactured 3D-printed OECTs have been explored and successfully demonstrated, this approach remains limited by relatively low resolution, typically ranging between 120 and 150 μm.[17] Here is where subtractive manufacturing can contribute to further improve the resolution. Based on the results of laser processing parameters and their effect on channel opening profiles, we selected a laser power of 500 mW at -50 μm interface with a 10× objective as the optimal conditions for patterning of horizontal OECT (*h*OECT) channels (**Figure 3a**). The whole fabrication process is outlined in **Figure S10**. The resulting *h*OECT channels have a width of 500 μm and minimum channel length of 1.25 μm (**Figure 3b**). Optimization of micrometer scale patterning can be observed in **Figure S11**.

Following laser patterning, the PEDOT:PSS dispersion was drop-casted onto the channel area (see Experimental Section for details). Output characteristics show that PEDOT:PSS *h*OECTs operate in hybrid mode for negative drain voltages and gate voltages ($V_G$) between -0.9 V and 1 V, as shown in **Figure 3b**. These results are consistent with previous PEDOT:PSS-based OECTs operating in depletion or hybrid mode depending on the channel width-to-length ratio and thickness.[32] Transfer characteristics measured at $V_D$ = -0.9 V reveal a threshold voltage ($V_{TH}$) of 0.8 ± 0.1 V and a maximum transconductance of 0.3 ± 0.1 mS at a $V_G$ = 0.1 ± 0.1 V (**Figure 3c** and representative gate currents in **Figure S12**). These results confirm that this laser patterning method can be successfully employed to fabricate *h*OECTs.

### 2.3. Selective cellulose acetate ablation for vertical OECTs

Vertically stacked OECT configurations, where the channel is sandwiched between the source and drain electrodes, have been introduced to reduce the channel length and enhance transistor density for circuit-level applications.[33] An insulation layer is typically used to prevent direct exposure of the electrode contacts to the electrolyte during electrical characterization.[34] Cicoria et al. demonstrated one of the first examples of printed vertical

OECT using a printed circuit board (PCB) printer.[34] However, incorporating the insulation layer introduces an additional challenge: creating defined pathways that allow the electrolyte to access the channel, which is essential for device operation.

Here, we employed a modified fabrication process to the one used for *h*OECTs to accommodate the vertical architecture, where the conducting polymer resides between the source and drain electrodes (**Figure 3d**). The fabrication process begins with printing and annealing the source electrode on the cellulose acetate substrate, as previously described. PEDOT:PSS is drop-casted directly onto the source contact and annealed (see Experimental Section, **Figure S10b**). A second graphene layer is then printed perpendicularly to the first electrode to form the drain electrode. The entire device stack is subsequently encapsulated with a spin-coated cellulose acetate layer for complete electrode insulation. To enable contact between the electrolyte and the channel material, we used the femtosecond laser to ablate an opening in the cellulose acetate just outside each of the four corners of the channel area of 125 μm vs 125 μm. These four openings give short diffusion distances for efficient electrochemical dope the *v*OECT channel.

This exposed the underlying PEDOT:PSS channel to the electrolyte without directly patterning the graphene, as illustrated in (**Figure 3d**). The channel length was determined by the thickness of the drop-cast PEDOT:PSS film (Dektak, experimental section).

Output characteristics indicate that the *v*OECT operates at similar drain and gate voltage ranges as the *h*OECT, but with higher maximum drain currents (**Figure 3e**), confirming that the electrolyte can indeed reach the PEDOT:PSS channel through the ablated areas. Transfer curves reveal that *v*OECTs exhibit a sharper ON/OFF transition at a more positive regime, with a threshold voltage shift to a more positive potential of $V_{TH} = 0.9 \pm 0.2$ V (**Figure 3f** and **S12**). These changes result in an increased maximum transconductance $g_m = 0.7 \pm 0.2$ mS at $V_G = 0.6 \pm 0.2$ V, which is consistent with shorter channel lengths compared to *h*OECTs[35] (525 nm with respect to 1 μm for *h*OECTs, see experimental section). Additionally, gate currents for *h*OECT and *v*OECT are reported in **Figure S12**.

## 2.4. Graphene-based gate electrodes and organic electrochemical inverters

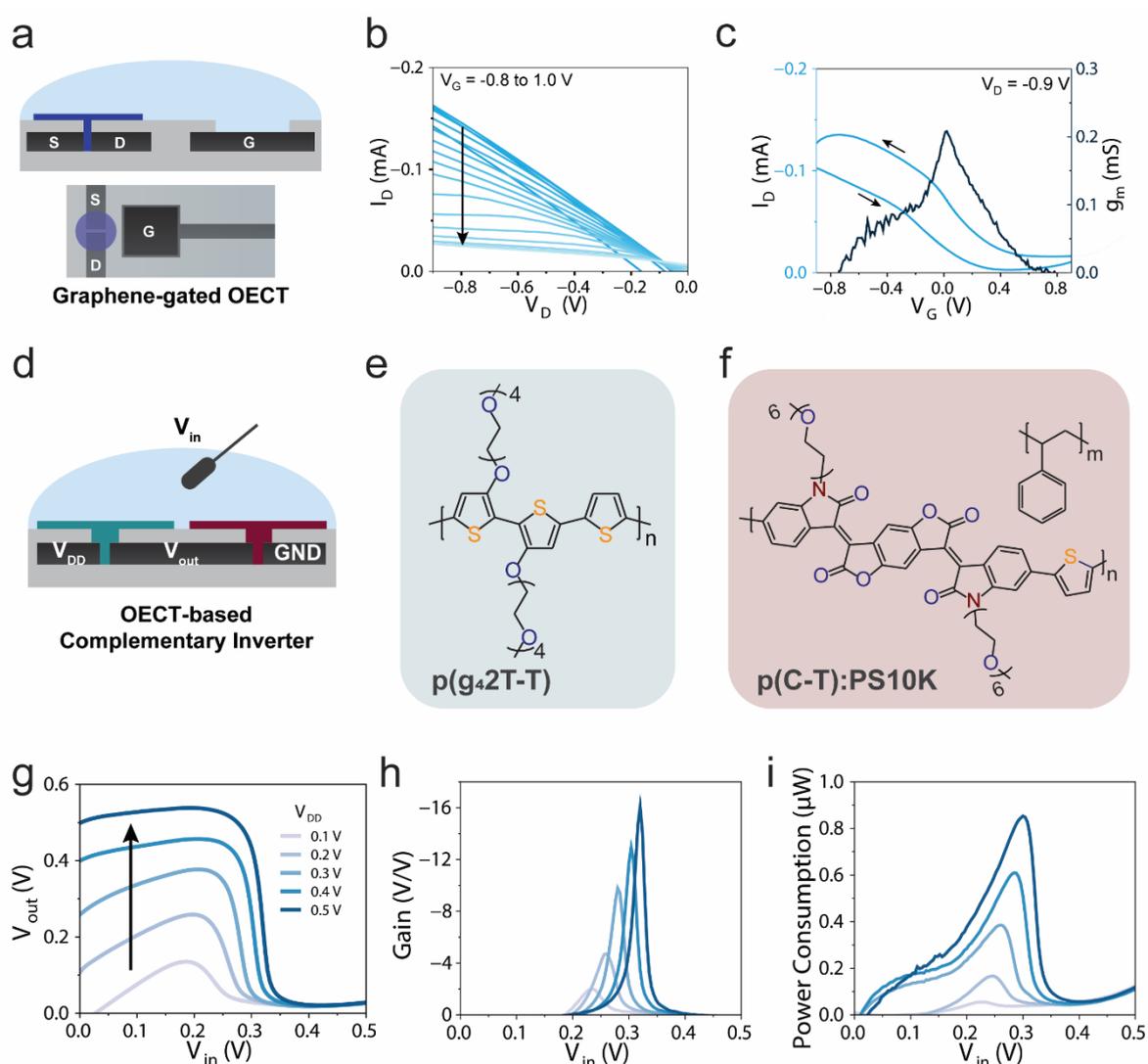

**Figure 4. Graphene-gated OECT and OECT-based complementary inverter.** (a) Schematic of the graphene-gated *h*OECT with an in-plane graphene gate electrode, represented with both (top) cross-sectional and (bottom) top view. (b) Output characteristics of the graphene-gated *h*OECT, with $V_D$ ranging from 0 to –1 V and $V_G$ from –0.9 to 1 V, with an inset from one of the devices fabricated with this method. (c) Transfer characteristics of the *h*OECT with respect to the gate voltage of planarly patterned graphene electrode. (*channel W = 500 μm, L = 1 μm, gate = 3000 × 3000 μm square*). (d) Schematic representation of the OECT-based complementary inverter circuit, where $V_{DD}$ is connected to the p-type OECT channel (p(g$_4$2T-T)), with $V_{out}$ subsequently leading to the n-type OECT channel (p(C-T)). Both channels are gated with $V_{in}$, which serves as the reference electrode. (e–f) Chemical structure of the p-type p(g$_4$2T-T) and n-type p(C-T). (g) Voltage transfer characteristic, (h) calculated voltage gain, and (i) power consumption characteristics of complementary inverter based on p(g$_4$2T-T) and p(C-T) interconnected p/n OECTs.

Developing a method to expose graphene electrodes without compromising their integrity is essential to fabricating planar OECTs with printed gates. Building upon the method of simultaneous laser ablation of graphene and cellulose acetate, we established a selective ablation strategy that effectively removes only the cellulose acetate encapsulation while preserving the underlying graphene electrodes. We adjusted the Z-height to shift the focal point from the graphene surface to the air/encapsulation layer interface. The laser power was reduced to 500 mW (using a 4× objective), below the graphene ablation threshold but sufficient to ablate cellulose acetate (**Figure 2f,g**). The process begins with an initial slow scan at 100 μm/s, followed by a fast pass at 3000 μm/s to remove residual polymer. We used the method both to open planar gates and to open contact pads (**Figure S13**).

We performed electrochemical analysis using a three-electrode setup on 3 × 3 mm (9 mm²) patterned graphene electrodes (**Figure S14**) to assess successful electrodes opening and stability in 1X PBS (aq.). The open-circuit potential (OCP) remained stable near 0 V for one hour, indicating reliable operation without delamination.[36] Cyclic voltammograms show quasi-rectangular shapes with minimal redox peaks, confirming electric double-layer capacitance.[37] Scan rate–dependent CVs (20–500 mV/s, 10 cycles) remain symmetric and reversible, exhibiting a proportional current–scan rate relationship. This behavior indicates stable, purely capacitive characteristics with minimal polarization.[38] Impedance spectroscopy reveals a relatively low impedance ($10^2$–$10^3$ Ω) from 1 Hz to 100 kHz at 0.0 V, indicating rapid charge transfer and minimal resistive losses. Based on these results, we tested in-plane graphene electrodes as planar gates for *h*OECTs.

Output and transfer characteristics show that the devices operate in a hybrid mode, similarly to *h*OECTs with identical channel dimensions (W = 500 μm and L = 1μm) with Ag/AgCl as the gate. The planar OECTs show a similar $V_{th}$ = 0.8 ± 0.3 V, along with increased hysteresis between the forward and reverse sweep, and a decrease in transconductance peak of $g_m$ = 0.2 ± 0.1 mS at higher $V_G$ = 0.2 ± 0.1 V, compared with the *h*OECT gate with Ag/AgCl. Such a decrease in performance with respect to devices having an Ag/AgCl-gate is attributed to the polarizable nature of the graphene gate in contrast to the non-polarizable behavior of Ag/AgCl.

To further demonstrate the versatility of our platform for integrating other conjugated polymers and circuit configurations, we developed complementary inverter circuits by integrating p-type and n-type OECTs (**Figure 4d**). The p-type OECT used p($g_4$2T-T) as the channel material, while the n-type counterpart was based on p(C-T):polystyrene 10 kDa (1:6) blend, selected for its improved operational stability with respect to pristine p(C-T) (**Figure 4f–** **g**).[39] Both materials were deposited by drop-casting on laser-patterned channels with W = 500 μm and L = 10 μm.

We measured voltage transfer characteristics of the inverter by sweeping the input voltage from 0 to 0.5 V under incremental supplying voltage ($V_{DD}$) steps of 0.1 V, ranging from 0.1 V to 0.5 V (**Figure 4g** and **Figure S15**). The voltage gain, defined as $\partial V_{out}/\partial V_{ini}$, reaches a maximum of 15 V/V at $V_{DD}$ = 0.5 V, with a switching threshold $V_M$ of 0.32 V (**Figure 4h**). These results are in line with previously reported gains for printed complementary OECT inverters[40]

Output currents reveal an asymmetric switching behavior, with bending at low $V_{IN}$, attributed to the operation of the p-type transistor, which does not fully turn off at zero gate voltages, unlike the n-type OECT (**Figure S15b**). Further optimizations, such as tuning the channel dimensions, can be used to improve the match in transient characteristics and further optimize the inverter behavior, but are beyond the scope of this study.[41]

We analyzed power consumption by extracting the inverter's current from its $V_{OUT}/V_{IN}$ characteristics (**Figure 4i**). Across all $V_{DD}$ values, the static power consumption remains below 0.86 μW operation achieved even at the supply voltage of 0.5 V, highlighting the circuit's suitability for low power applications. The low-voltage operation and clear inversion demonstrates the platform´s compatibility with a variety of organic mixed ionic/electronic conductors and include more complex transistor configurations to create logic gates with cleanroom-free fabricated flexible devices.

## 2.5. Thermal degradation and sustainable device lifecycle

Adopting carbon-based materials such as graphene and cellulose acetate offers a promising route toward more environmentally sustainable OECTs, addressing concerns associated with conventional inorganic components like gold and chromium. In contrast to typical insulating polymers used for device passivation - such as Parylene C, which requires chemical vapor deposition - cellulose acetate can be deposited via a simple room-temperature spin-coating process, reducing energy consumption and fabrication complexity. Similarly, extrusion printing relies on inexpensive equipment and nozzles and enables printing of water-based inks at room temperature.

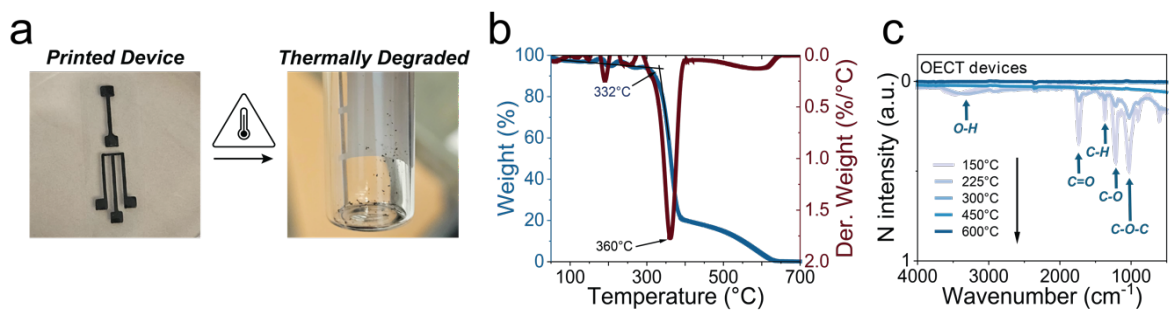

**Figure 5.** Thermal degradation of all printed *h*-OECTs. (a) Before (left) and after (right) thermal degradation. (b) TGA results and critical temperatures to consider, scanning from room temperature to 700°C. (c) FTIR for fully fabricated *h*-OECT devices at different temperatures from 150°C to 600°C.

To evaluate the environmental footprint of our all-carbon OECTs, we investigated the thermal degradation behavior of devices made of graphene and cellulose acetate (**Figure 5**). Traditional OECTs often incorporate metals such as gold, silver, chromium, and titanium, which demand energy-intensive processing due to their high melting points (e.g., gold at 1064 °C, silver at 961 °C, chromium at 1907 °C, and titanium at 1668 °C) and environmentally hazardous recycling methods, such as acid leaching.[42] In contrast, carbon-based devices can be degraded by thermal processes at significantly lower temperatures.

Thermogravimetric analysis (TGA) of our devices, composed of graphene and cellulose acetate, show degradation occurring around 360 °C using air in a standard laboratory oven. The thermal decomposition profile up to 700 °C (**Figure 5b**) reveals an initial decomposition onset at 332 °C, followed by a major degradation peak near 360°C. This process results in minimal residue (~0.2% ash with by weight), indicating that 80% of the device mass undergoes thermal decomposition at 360°C.

To further investigate the chemical changes underlying this degradation, we employed Fourier-transform infrared (FT-IR) spectroscopy to analyze the thermal breakdown of cellulose acetate within the *h*OECTs. The FT-IR spectra revealed a significant reduction in characteristic functional groups above 300 °C, including O-H stretching vibrations (3400–3200 cm$^{-1}$), aliphatic C-H stretching (3000–2800 cm$^{-1}$), C=O stretching (1750 cm$^{-1}$), and C-O stretching and C-H bending vibrations (1500–1000 cm$^{-1}$). We observed similar changes in cellulose acetate controls (**Figure S16**), confirming that these spectral changes can be attributed to the degradation of ester, hydroxyl, and aliphatic hydrocarbon groups of cellulose acetate.[22] The results are consistent with other works, identifying $H_2O$, $CO_2$, CO, and nonvolatile carbonaceous residues as the main products of cellulose thermal decomposition,[43] which represent approximately 95% of the total volume per device alongside graphene and conducting polymers.

Given that most disposable devices are ultimately incinerated with municipal solid waste streams,[44] our results support the feasibility of using low-energy thermal degradation as an end-of-life strategy. This highlights the potential of degradable carbon-based electronics for single-use applications in bioelectronics.

## 3. Conclusion

In this work, we presented a scalable, lithography-free microfabrication strategy for fully organic, self-standing, metal-free, carbon-based, flexible OECTs, leveraging a hybrid additive-subtractive approach. By combining spin-coating of cellulose acetate substrates, extrusion printing of aqueous graphene ink, and femtosecond laser ablation of cellulose acetate insulation layers, we achieved micron-scale resolution and cleanroom-free device production using degradable, carbon-based materials.

We validated the versatility of this approach through the fabrication of multiple OECT architectures, including horizontal, vertical, and planar-gated configurations, with channel lengths down to 1 μm and electrode sheet resistances below 10 Ω/sq. This platform supports the integration of functional circuit elements, such as complementary inverters based on p-type and n-type OECTs, and enables the realization of fully patterned, flexible devices with printed graphene gate electrodes. Thermal degradation analysis shows that over 80% of the device mass decomposes below 360°C, offering a low-energy end-of-life strategy compared to conventional metal-based electronics.

Overall, this work bridges sustainable material development with advanced microfabrication, providing a practical route towards the development of high-performance, low-footprint organic bioelectronic devices. By combining high-resolution patterning, environmentally friendly materials, and scalable manufacturing, this approach lays the foundation for next-generation flexible OECT-based systems in wearable, implantable, and disposable sensing applications.

## 4. Experimental Section

Graphene ink (7 wt% in water, product no. 805556), cellulose acetate (Mn ~50,000, product no. 419028), and poly(vinyl alcohol) (PVA, Mw 31,000-50,000, 98-99% hydrolyzed, product no. 363138) were purchased from Sigma-Aldrich. Poly(3,4-ethylenedioxythiophene) doped with poly(styrene sulfonate) (PEDOT:PSS, PH 1000) was obtained from Heraeus (Clevios™). Additional chemicals—DBSA (4-dodecylbenzenesulfonic acid, product no. 44198), GOPS ((3-

glycidyloxypropyl)trimethoxysilane, product no. 440167), and ethylene glycol (EG, product no. 324558)—were also sourced from Sigma-Aldrich. The n-type semiconductor p(C-T) (3,7-dihydrobenzo[1,2-b:4,5-b']difuran-2,6-dione) was synthesized according to previously reported literature[45] and diluted with 10 kDA polystyrene (1:6 monomer ratio) to improve long-term stability[46]. The p-type semiconductor p(g$_4$2T-T) (poly(3,3'-bis(tetraethylene glycol methyl)-2,2'-dithiophene-thiophene)) was also prepared based on previously reported protocols[47].

*Formulation of inks for substrates and insulation:* a 17.5 wt% polyvinyl alcohol (PVA) solution was prepared by dissolving 42.4 g of PVA in 200 mL of deionized water at 60°C with magnetic stirring. The solution was stirred until homogeneous and left at room temperature for 24 hours to achieve full clarity. For the cellulose acetate solution, 4.5 g of cellulose acetate was prepared in a 50 mL Falcon tube and dissolved using 45 mL of acetone by manual agitation and left fully dissolve overnight at room temperature.

*Preparation of the device substrate:* a 6-inch glass wafer was sequentially washed with soap, MiliQ water, and then soaked in acetone and isopropanol for 5 minutes for each step, then dried with air. The cleaned wafer was spin-coated with the PVA solution (8 mL, 1000 RPM, 120 seconds), then placed on a hot plate at 80°C for 20 minutes to dry the excess of humidity. The cellulose acetate solution of 8 mL was subsequently spin-coated at 500- 2500 RPM for 120 seconds to obtain films of different thicknesses. The acetone solvent evaporated naturally at ambient temperature during the spin-coating process, forming a solid, transparent thin cellulose acetate film.

*Fabrication of planar OECT*: all the graphene electrodes were fabricated via direct ink writing using FELIX BIOprinter (FELIXprinters, Netherlands) using a 5 mL Omnifix® syringe and nozzles (Diatom A/S) with a diameter of 0.20 mm. Electrode patterns were generated using custom-written G-code in Repetier opensource software. Prior to printing, the cellulose acetate substrate was air plasma-treated for five minutes using a handheld corona surface treater (Aurora Pro Scientific). The substrate was then transferred to the bed of the printer and the graphene ink was loaded into the syringe. After printing, the printed graphene electrodes were annealed at 130 °C for 10 minutes. A second layer of cellulose acetate was then spin-coated on top to encapsulate the electrodes at 1500 RPM for 120 sec to passivate the electrode contacts. PEDOT:PSS ink (containing PEDOT:PSS PH 1000 V/V 92.5%, GOPS V/V 1%, EG V/V 6%, and DBSA V/V 0.5% as additives) was diluted 1:4 with Milli-Q water. This formulation was then drop-cast between the femtosecond laser patterned drain-source channels. Afterwards, the PEDOT:PSS film was annealed at 120 °C for 20 minutes.

*Fabrication of vertical OECTs:* for vertical devices, cellulose acetate substrate and glass wafer preparation prior to graphene printing was performed similar to planar OECT fabrication. After the printing of first graphene layer, the electrodes were dried at 130 °C for 10 minutes. 0.2 µL of PEDOT:PSS ink (containing PEDOT:PSS PH 1000 V/V 92.5%, GOPS V/V 1%, EG V/V 6%, and DBSA V/V 0.5% as additives) was diluted 1:4 with Milli-Q water, and drop-casted between source and drain electrodes and dried at 120 °C. Afterward, a second layer of graphene was printed perpendicularly to the previous graphene layer and was dried at 130 °C for 10 min. Finally, a second layer of cellulose acetate was spin coated at 1500 RPM for 120 sec to encapsulate the device prior to femtosecond ablation on the four corners of the intersection of the *v*OECT channel avoiding to ablate graphene electrodes.

*Fabrication of organic complementary inverter:* OECT-based inverters used p-type polymer p(g$_4$2T-T) dissolved in chloroform (5 mg/mL) and n-type polymer p(C-T) blended with 10 kDa polystyrene (1:6 monomer ratio)[46] in chloroform (concentrations?). Both materials were deposited as channel materials via drop-casting of 0.2 µL solution followed by annealing at 120°C for 20 min. An Ag/AgCl pellet was used to gate both transistors and control the input voltage ($V_{IN}$), with PBS 1X as the electrolyte. The inverters were designed using channel geometry of W = 500 µm and L = 10 µm, with two interconnected channels.

*Femtosecond laser patterning:* the subtractive patterning was performed using a femtosecond laser workstation equipped with a laser source (Spirit 1040-4-SHG, Spectra-Physics, USA) and a linear motorized stage (XMS100, Newport, USA) to create the lateral opening for graphene and cellulose acetate. The sample was irradiated using 4× (Plan Achromat RMS4X, Olympus, Japan), 10× (Plan Achromat RMS10X, Olympus, Japan), and 20× (Plan Achromat RMS20X, Olympus, Japan) microscope objective to focus the laser source at different beam spot size. The topmost surface interface was set by performing low-power laser irradiation, assisted by a three-dimensional ruler calibration design to define the 0 µm Z-height interface. The laser source was set to expose 520 nm laser pulse at 298 fs duration for each single pulse, where the typical working range to achieve ablation patterning occurred at 1MHz laser pulse frequency, laser pulse energy between 300-1000 mW, and scan speed 10-3000 µm. Successful patterning was confirmed by upright microscope (Axiolab 5, Carl Zeiss, Germany) observation with transmitted light illumination setting to verify the patterned features.

*Electrochemical and electrical device characterization:* OECT and complementary amplifier devices were characterized using a Keithley 4200A-SCS parameter analyzer (Tektronix, USA), with two source measurement units (SMUs) and one pulse measurement unit (PMU). OECTs and inverter circuits were operated using PBS 1X (aqueous) solution as the electrolyte, with a silver/silver-chloride (Ag/AgCl) pellet serving as the gate electrode. A Keithley 2410 source meter was connected to the parameter analyzer to supply the $V_{DD}$ potential. Electrochemical characterization was performed using a BioLogic potentiostat with an impedance module. The working electrode was a printed graphene device, the pseudo-reference electrode was an Ag/AgCl pellet, and the counter electrode was a platinum electrode.

*Four-point probe measurements:* sheet resistance of the thin polymer films was measured using the Ossila Four-Point Probe system with spring-loaded, rounded, gold-plated probes to ensure precise and non-destructive surface measurements. The system provides real-time current and voltage readings across the probes and incorporates both positive and negative polarity measurements to enhance accuracy. Measurements were conducted using the integrated Ossila SMU, with results calculated using the proprietary software to account for geometrical correction factors.

*Profilometry and feature analysis:* the thicknesses of PEDOT:PSS drop-cast films were measured using a Dektak contact profilometer and found to be $525 \pm 78$ nm (average of five samples). The vertical and lateral surface height profile of the patterned sample was obtained using 3D optical profilometer based on coherence scanning interferometry (CSI) (Nexview NX2, Zygo Corporation, USA) at 10× objective at 2× zoom to perform stitched measurement (FOV: 420 μm; vertical res: 4 nm; lateral res: 0.410 μm). The raw data were subsequently processed with a colormap to visualize surface height variations and feature recognition to calculate surface parameters using custom Python scripts. The XY coordinates and Z-values data were used to perform regional analysis and provide graphene land CA lateral opening, as well as CA depth to the graphene topmost surface (see **Note S2** and **Figure S4**, Supporting Information).

*Raman spectroscopy and microscopy:* the graphene inks were characterized using Raman spectroscopy on a LabRAM HR 800 Raman instrument, equipped with a 600 grating and an air-cooled double-frequency Nd:YAG laser (532 nm, 50 mW). Following femtosecond laser ablation on the device, Raman spectroscopy and spatial spectral imaging were performed on confocal Raman imaging system alpha300 R (WITec GmbH, Germany), using 50× objective (Zeiss LD EC Epiplan-Neofluar Dic 50x/0.55) connected to UHTS300 spectrometer. Data acquisition was conducted through a 532 nm Nd:YAG laser source with laser power of 30 mW and 0.05 second exposure time per point, with 50×50 raster scans over 50×50 μm to 200×200 μm designated areas for spatial imaging. Acquired spectral data was then processed and evaluated using proprietary WITec Project SIX software to conduct cosmic ray removal (filter size 7, dynamic factor 8), background subtraction (polynomial fit order 6), and TrueComponent analysis to produce the representative average spectra and intensity spatial information of each component.

*Thermogravimetric analysis (TGA):* The devices were characterized using a TA Instruments Discovery, at a heating rate of 10°C min$^{-1}$ under an air flow rate of 20 mL.min$^{-1}$. The samples were place on a platinum HT Pan (Part number 957571.901 from TA instruments). Reduction of noise was performed on the TGA raw signal (below 310°C) to enhance data clarity. For this, the OriginPro's built-in Savitzky–Golay method was used with a polynomial of second order.

*Fourier Transform Infrared Spectroscopy (FTIR):* Several devices were placed onto an alumina crucible in a pre-warmed muffle furnace at various temperature (150°C, 225°C, 300°C, 450°C, and 600°C) for 60 minutes. The sample were subsequently analyzed using a Varian 670-IR Spectrometer equipped with an attenuated total reflectance accessory using a deuterated triglycine sulfate detector (ATR-DTGS). Spectra were recorded in absorbance mode in the range 4000-390 cm$^{-1}$ with a spectral resolution of 4 cm$^{-1}$ and 32 scans averaged per sample.

**Acknowledgements**

A.E.A.R. and J.J. contributed equally to this work. This work was in part financially supported by Digital Futures. E.Z. and A.E.A.R. gratefully acknowledge the Göran Gustafsson Foundation. E.Z. gratefully acknowledges the Wallenberg Initiative Materials Science for Sustainability (WISE) funded by the Knut and Alice Wallenberg Foundation, the Swedish Research Council (Grant No. 2022-02855), and FORMAS – a Swedish Research Council for Sustainable Development (Grant No. 2022–00374) for support. This work was supported by AIMES – The center for integrated medical and engineering sciences (www.aimes.se), Karolinska Institutet (1–249/2019), KTH Royal Institute of Technology (VF-2019-0110), and Getinge AB (4.1599/2018). A.E.A.R. extends special thanks to Sebastian Buchman for his invaluable assistance with the inverter setup for characterization, as well as for his mentoring, also to Yunfan Lin for the many fruitful discussions on OMIECs and microfabrication that contributed to the development of this project. We would also like to thank Alessandro Enrico for providing the three-dimensional ruler calibration design and for his support during proof-of-concept experiments for laser ablation. Additionally, we express our gratitude to Asaminew Y. Shimolo for providing p-type conducting polymers for preliminary testing of complementary inverters.

## Conflict of interest

The authors declare no conflict of interest.

**Supporting Information**

Supporting Information is available from *npj Flexible Electronics* Online Library or from the author.

**Carbon-based Microfabricated Organic Electrochemical Transistors Enabled by Printing and Laser Ablation**

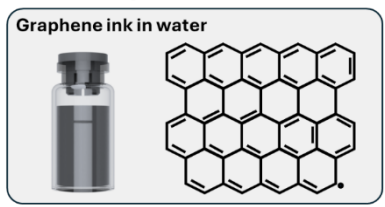
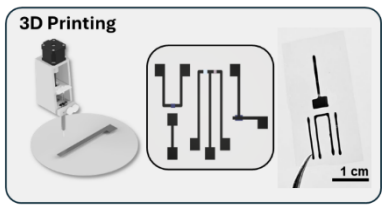
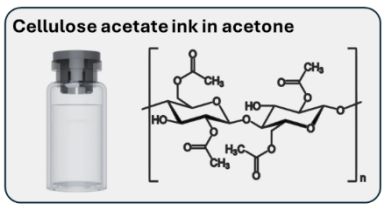
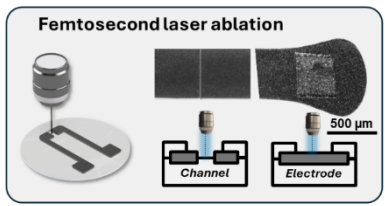

An additive–subtractive strategy enables the microfabrication of organic electrochemical transistors (OECTs) by combining graphene ink extrusion printing for electrodes with femtosecond laser ablation of cellulose acetate to define micrometer-scale channels. This approach enables rapid prototyping of various OECT architectures, including horizontal, vertical, and planar configurations, as well as a p/n-type NOT inverter, using biodegradable materials for sustainable organic electronics.

Supporting Information

**Carbon-based Microfabricated Organic Electrochemical Transistors Enabled by Printing and Laser Ablation**

*Alan Eduardo Avila Ramirez,[§] Jessika Jessika,[§] Yujie Fu, Gabriel Gyllensting, Marine Batista, David Hijman, Jyoti Shakya, Yazhou Wang, Wan Yue, Renee Kroon, Jiantong Li, Mahiar Max Hamedi, Anna Herland,\* Erica Zeglio\**

**Supplementary Video 1. Simultaneous ablation of graphene and cellulose acetate using 4× objective.** (mp4 file format).

This video demonstrates real-time femtosecond laser ablation using three passes. As the laser scans over the device, changes in light reflection reveal the formation of the ablated channel. Light angle was adjusted mid-video to enhance visual confirmation of opening depth and uniformity. This method removes both layers efficiently and is used mainly in horizontal OECT fabrication.

**Supplementary Video 2. Simultaneous ablation with 10× objective at sub-threshold power.** (mp4 file format).

The video begins with two passes of sub-threshold power. The second pass partially opens the graphene, and subtle changes in brightness and contrast (adjusted live) help determine whether full ablation is achieved. The third pass shows minimal further expansion, confirming that once the ablation threshold is reached, additional passes have limited effect on lateral dimensions.

**Supplementary Video 3. Simultaneous ablation with 20× objective at sub-threshold power.** (mp4 file format).

Here, more than 10 sequential passes with sub-threshold power are applied using 20× objective. Gradual material removal is observed, leading to complete graphene opening without significant change in lateral opening size. At the end, Z-height is adjusted to confirm complete penetration through the graphene layer.

**Supplementary Video 4. Selective ablation of cellulose acetate using 4× objective.** (mp4 file format).

This video shows progressive ablation of the cellulose acetate encapsulation layer without disturbing the underlying graphene. Initially, oxidation appears as a 'burn mark' (darkening), which starts diminishing on the second pass and becomes brighter on the third, indicating transition to the melting phase. This gradual shift reflects controlled thermal degradation of the cellulose acetate surface, used to expose vertical OECT channels without compromising electrode integrity.

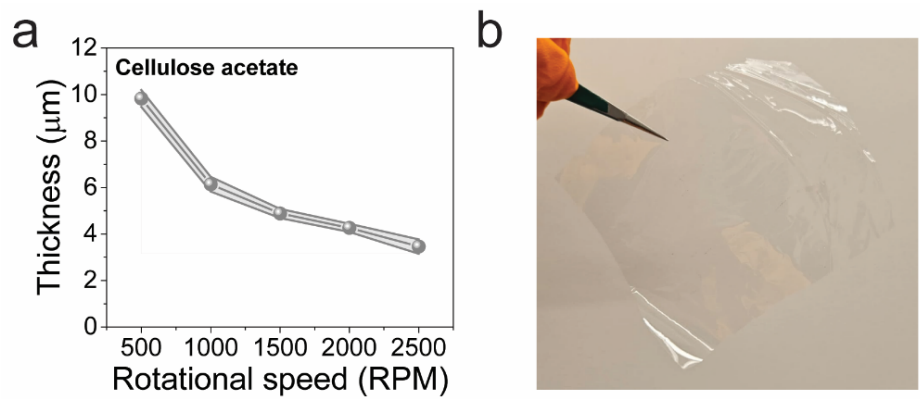

**Figure S1.** a) Thickness measurements of the spin-coated cellulose acetate substrate obtained using Dektak profilometer at different revolutions per minute to show the flexibility to change thickness with our developed ink. b) Photograph of the cellulose acetate film (~5 μm thick) after delamination from the carrier substrate.

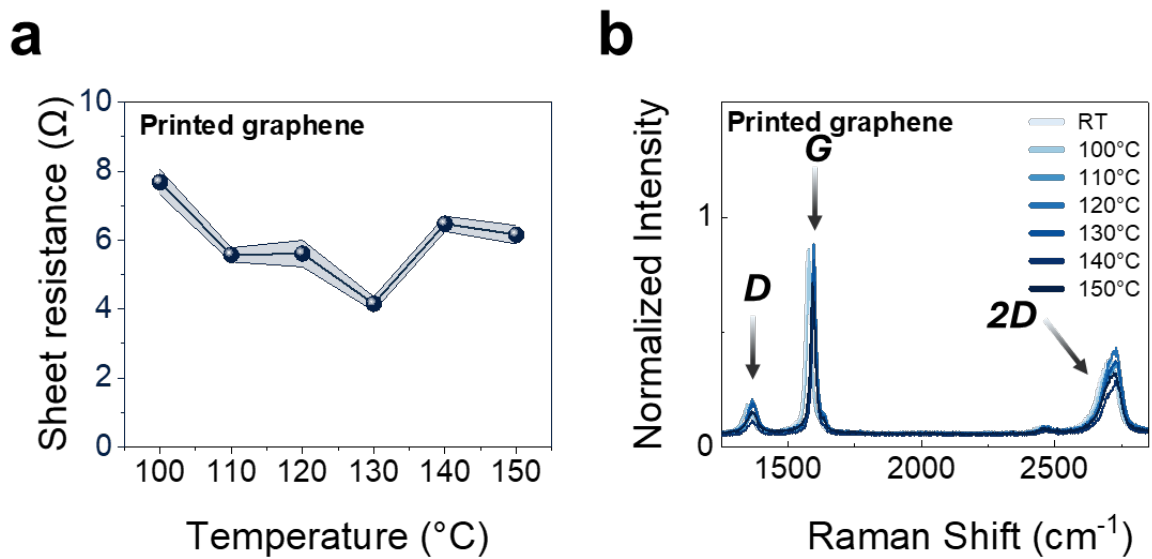

**Figure S2.** a) Four-point probe sheet resistance measurements of annealed graphene ink at different temperatures. b) Raman spectra of printed graphene patterns at different annealing temperatures. The D peak, located at approximately 1350 cm$^{-1}$, is due to the disordered structure of graphene.[1–3] The G peak, located at approximately 1600 cm$^{-1}$, arises from the stretching of the C-C bond of the sp$^2$ hybridized carbons of graphene. The 2D peak, found around 2717 cm$^{-1}$, results from second-order Raman scattering by in-plane transverse optical phonons near the boundary of the Brillouin zone of graphene. The full width at half maximum (FWHM) of the 2D peak for the graphene ink is measured at approximately 80 cm$^{-1}$, confirming the presence of multilayer graphene.

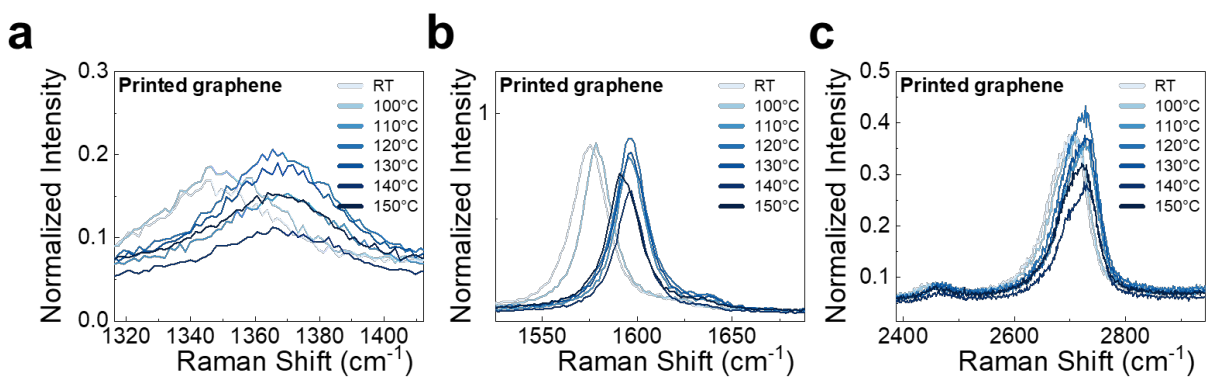

**Figure S3.** Zoom-ins of the Raman spectra showing the temperature dependance of the position of the a) D, b) G, and c) 2D peaks of printed graphene from 100 °C to 150 °C.

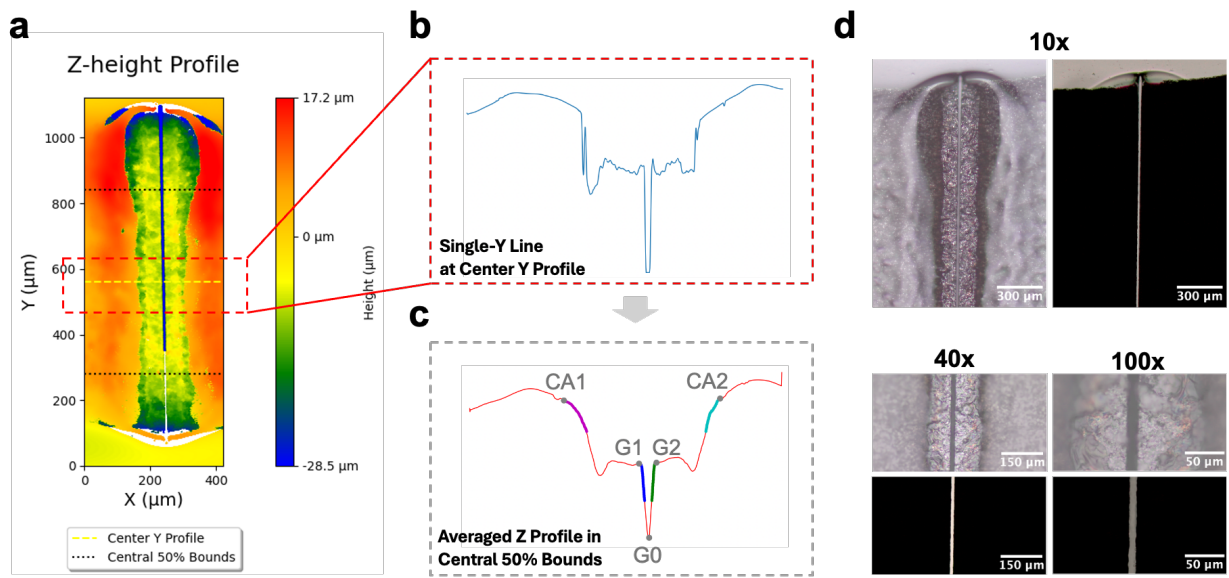

**Figure S4. 3D optical profilometry and automated feature recognition on patterned sample.** (a) Z-height profile colormap of the ablated sample surface, captured using a 3D optical profilometer. The yellow dashed line marks the single-line central Y-profile, while the blue dotted lines indicate the central 50% region for the lateral analysis. (b, c) Extracted Z-height profiles from the central single-line Y-profile and averaged region with key feature points annotated. G1 and G2 represent local maxima on the graphene surface, while G0 is the global minimum corresponding to the deepest point. CA1 and CA2 represent the lateral opening of the cellulose acetate (CA) insulator layer, defined by the first derivative minima (left) and maxima (right). (d) Further verification of successful graphene opening was performed via optical microscopy, showing both the top view of and the transmitted light illumination view of the ablated regions at 10x, 40x, and 100x magnifications to verify graphene exposure and channel clarity.

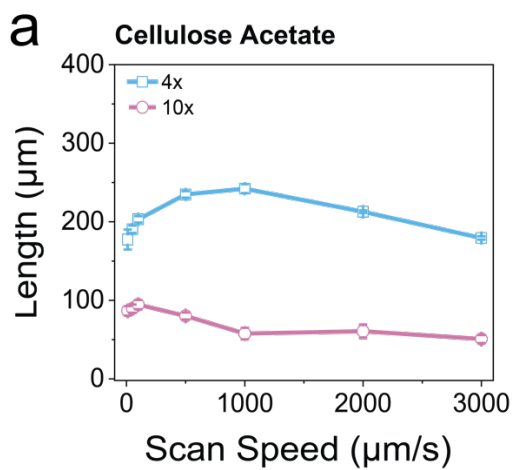
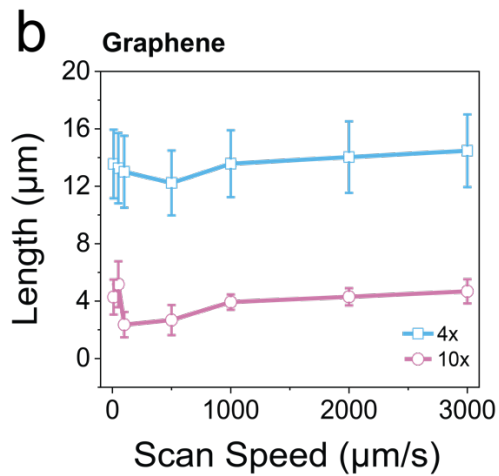

**Figure S5. Influence of speed in laser ablation on lateral opening length.** In (a) graphene, and (b) cellulose acetate.

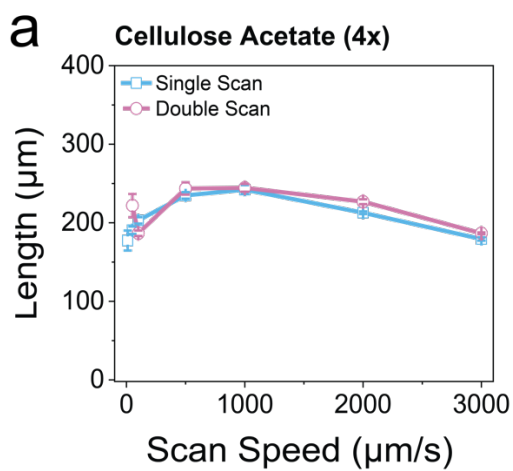
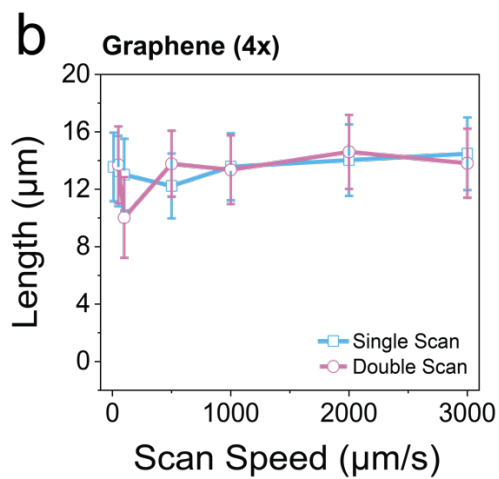

**Figure S6. Influence of pass number on lateral opening length via laser ablation with 4x objective.** In (a) graphene, and (b) cellulose acetate.

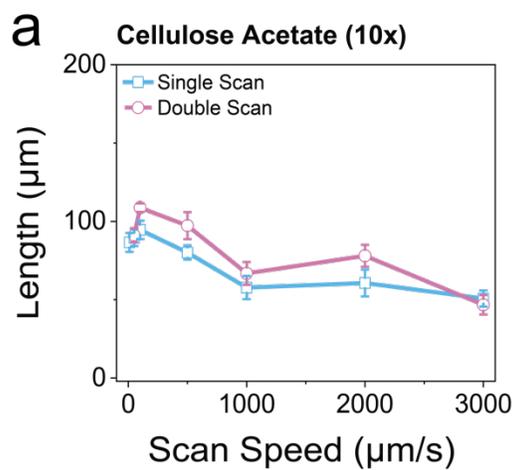
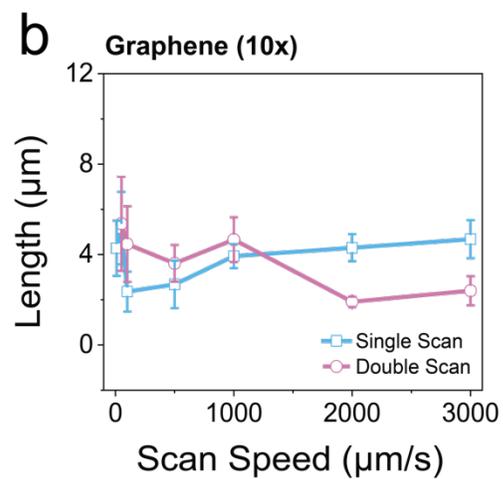

**Figure S7. Influence of pass number on lateral opening length via laser ablation with 10x objective.** In (a) graphene, and (b) cellulose acetate.

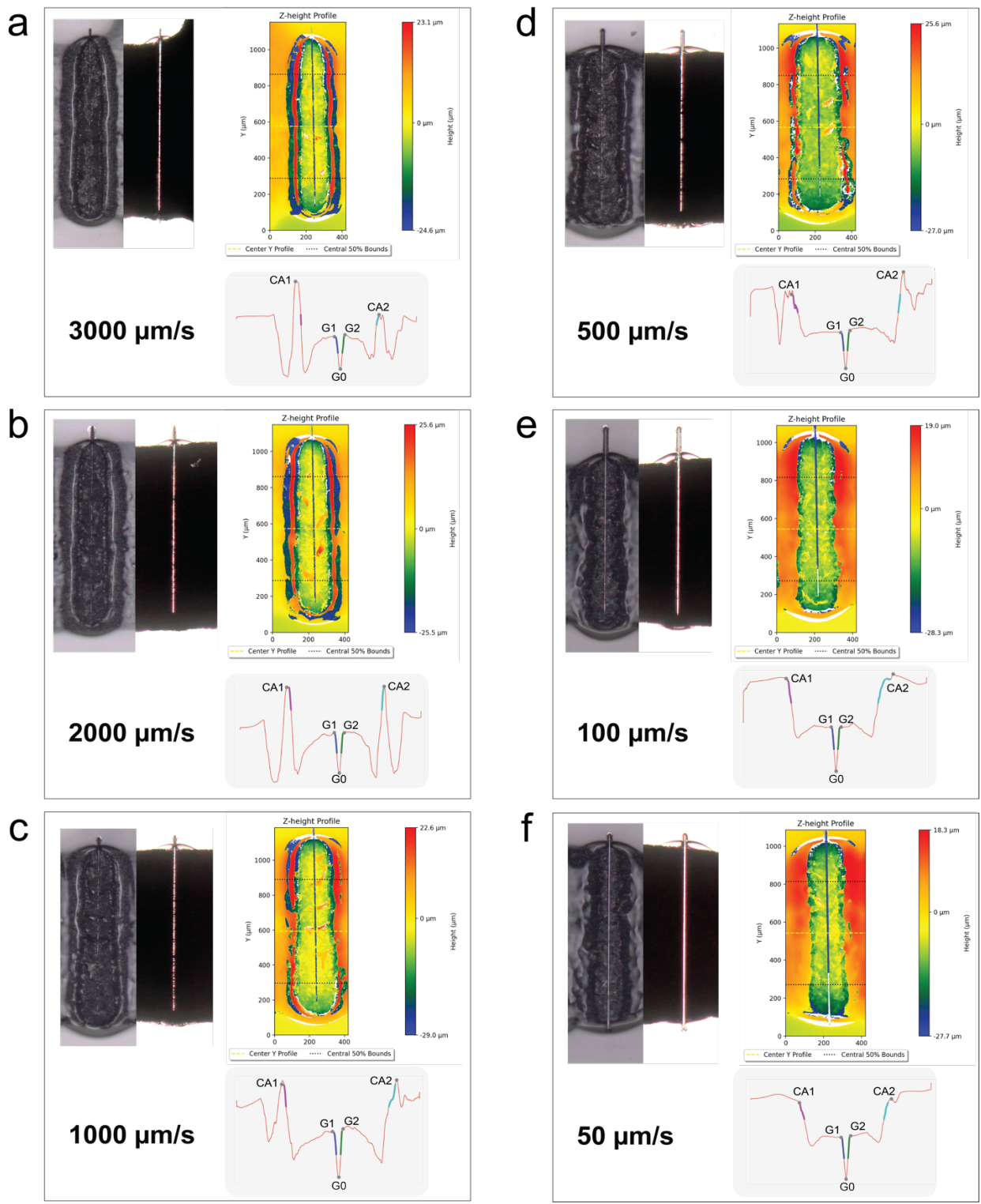

**Figure S8. Effect of scanning speed on graphene and cellulose acetate features with 4x objective.** A series of images and analyses demonstrating the effect of scanning speed on graphene and cellulose acetate lateral openings, performed with 4x objective at 1000 mW and -150 μm Z-height. Each panel shows optical microscopy images of the ablated regions, with reflecting (RL)

and transmitting light (TL) views, for scanning speeds ranging from 3000 µm/s to 50 µm/s. The Z-height profiles obtained using 3D optical profilometer are represented with the colormap indicating height variations and the extracted height profiles from the averaged 50% central Y bounds region, with key feature points for cellulose acetate and graphene annotated as CA1, CA2, G0, G1, and G2.

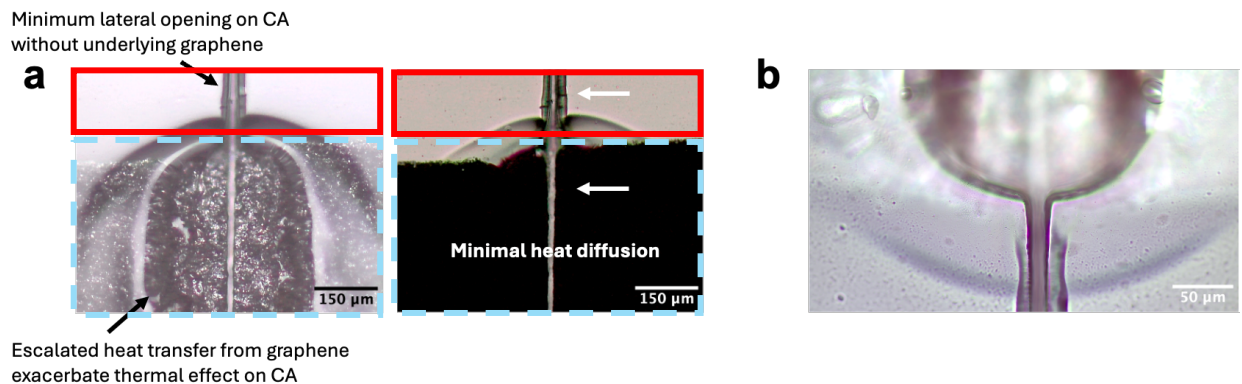

**Figure S9. Graphene-enhanced thermal ablation of cellulose acetate.** (a) Reflecting (left) and transmitting (right) microscope images showing the ablation of the cellulose acetate layer. The presence of underlying graphene significantly enhances thermal diffusion, resulting in a wider lateral opening of cellulose acetate (highlighted in light blue, dashed line). In contrast, regions without graphene underneath (highlighted in red) exhibit limited heat diffusion, leading to a more confined ablation profile. Notably, the lateral opening in these regions closely matches that of the graphene layer, as seen in the right image. (b) Optical microscope image of the cellulose acetate surface after ablation, illustrating how the enhanced thermal conductivity of the underlying graphene increases the efficiency of cellulose acetate ablation through localized heat transfer.

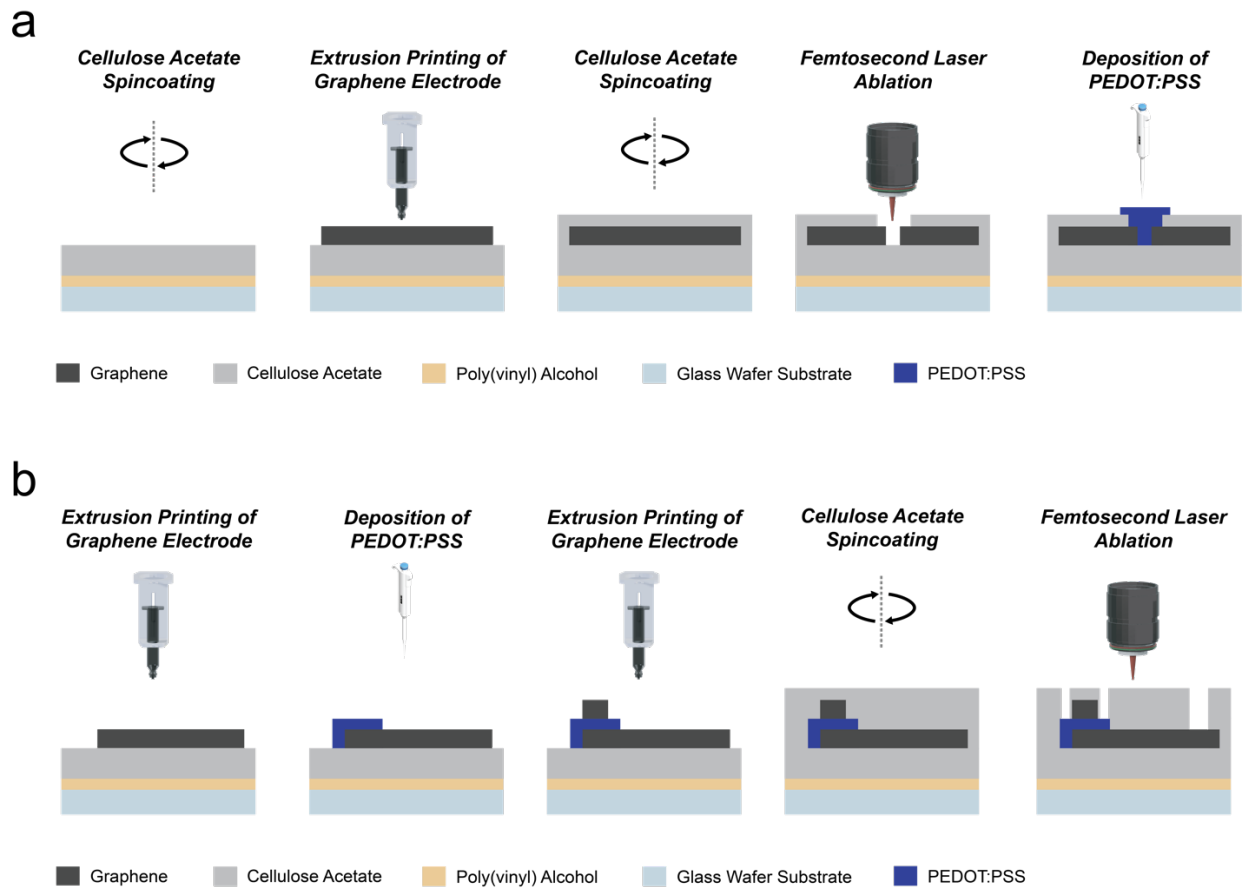

**Figure S10.** (a) Flowchart schematic from the fabrication of hOECT. (b) Flowchart schematic from the fabrication of vOECT.

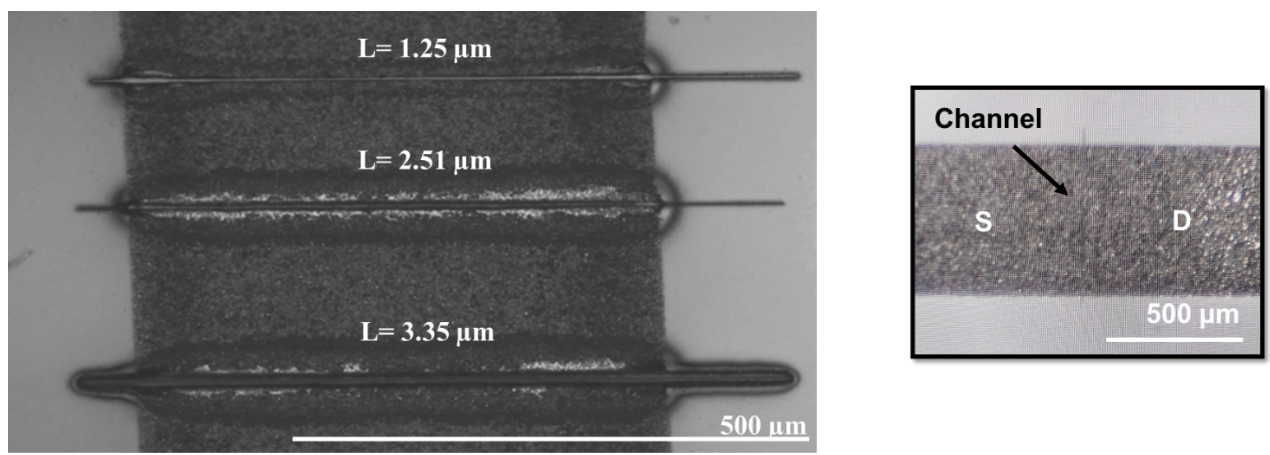

**Figure S11.** Examples from the downscaling of the ablation in OECT patterning. The figure on the left shows examples of the minimum achieved resolution. The example on the right

corresponds to an *h*OECT channel prior to the coating of the conducting polymer, shown at a 1 µm scale length opening.

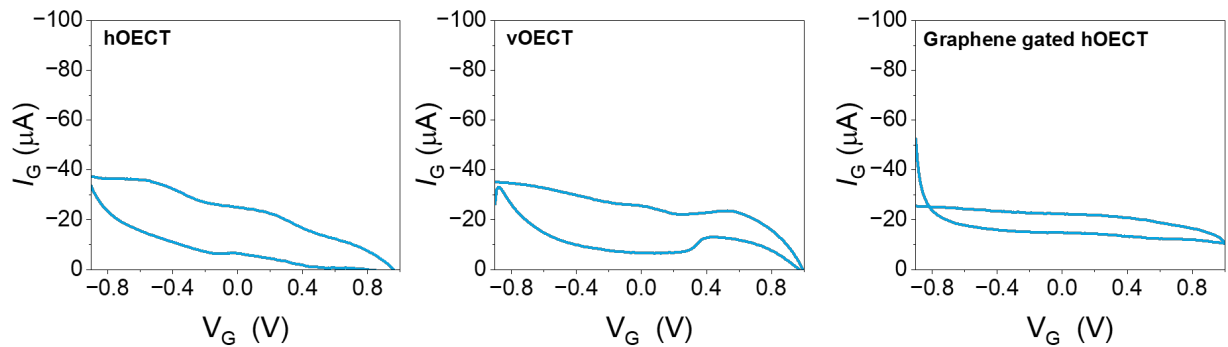

**Figure S12** Gate current from transfer characteristics (a) hOECT with Ag/AgCl gate electrode; (b) vOECT with Ag/AgCl gate electrode; (c) hOECT with gate electrode graphene 3 x 3 mm.

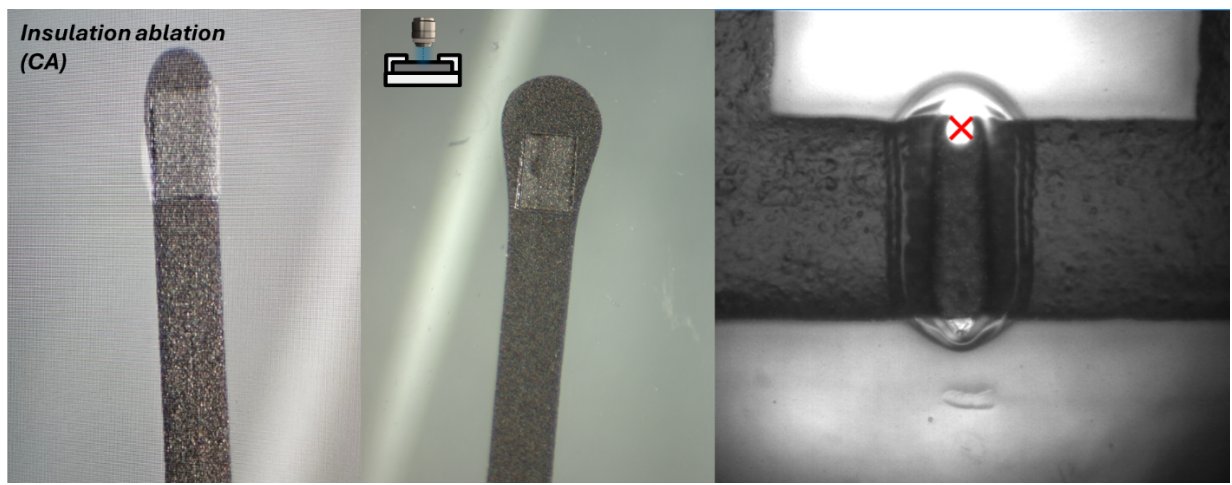

**Figure S13.** Representative examples of contact pad openings achieved via selective laser ablation of the cellulose acetate layer, with no observable damage to the underlying printed graphene structures. The third image on the right illustrates the laser in operation, precisely removing the cellulose acetate layer while preserving the integrity of the graphene electrode beneath.

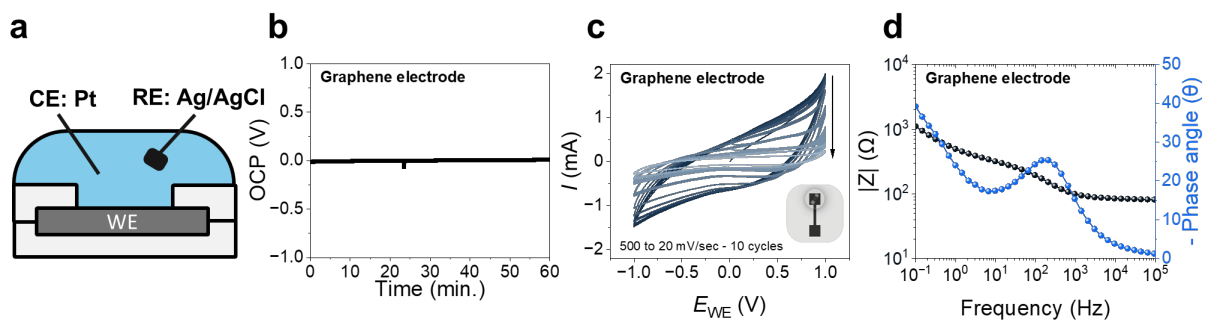

**Figure S14.** (a) Schematic of a single-graphene electrode after cellulose acetate ablation, used as the working electrode (WE) in a three-electrode setup. (b) Electrochemical stability assessed through cyclic voltammetry at scan rates ranging from 500 mV/s to 20 mV/s, with 10 cycles for each scan rate. (c) Impedance spectroscopy analysis of the electrode, showing low-impedance regimes and capacitive behavior at 0 V.

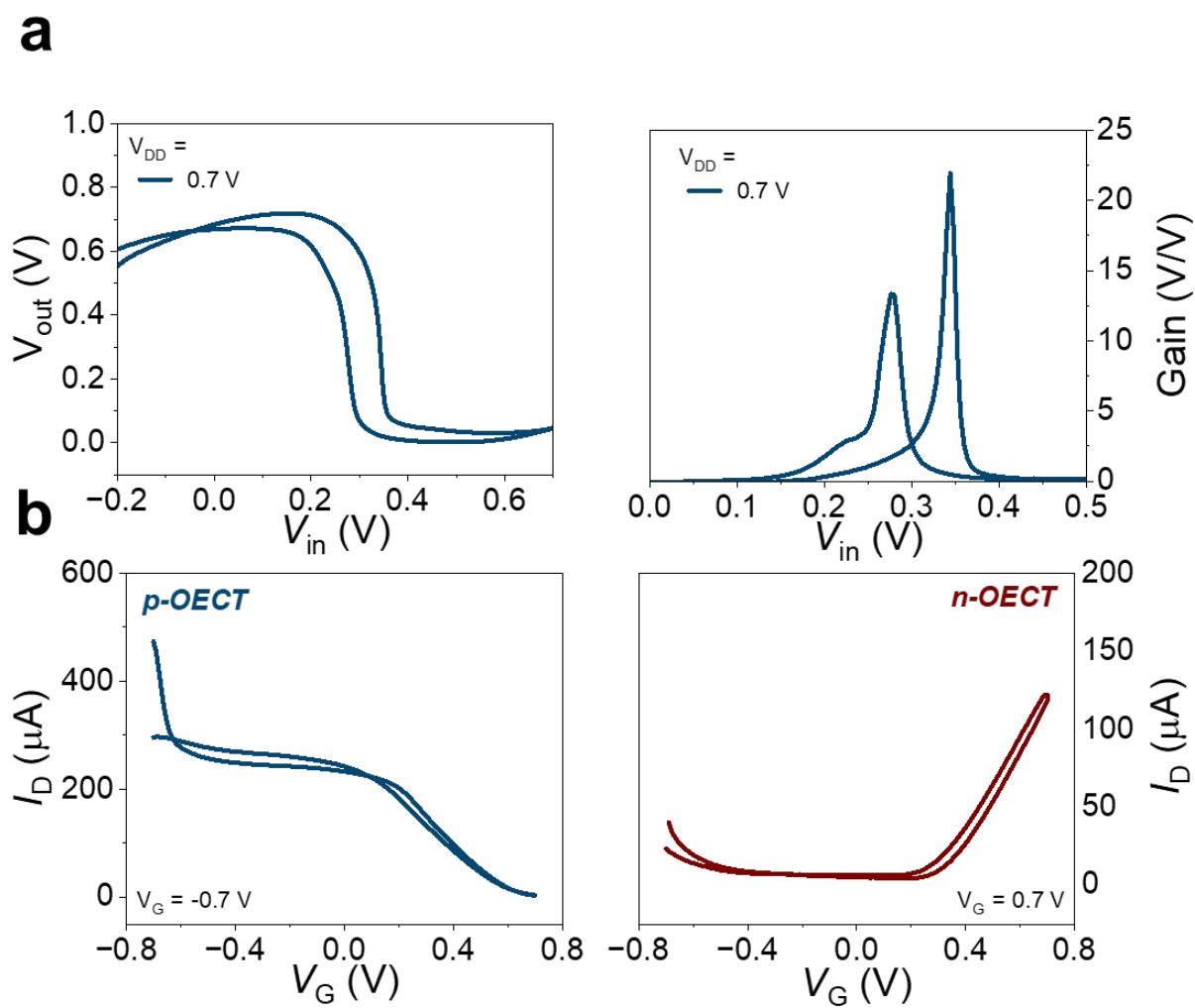

**Figure S15.** (a) Organic complementary inverter operating at a $V_{DD}$ of 0.7 V, above saturation in the initial cycle, resulting in a gain of 22 V/V, comparable to literature values. (b) Transfer characteristics of p(g$_4$2T-T) as the p-type OECT and p(C-T) as the n-type OECT. This represents the highest $V_{DD}$ and $V_{in}$ applied to these devices to avoid degradation at higher potentials.

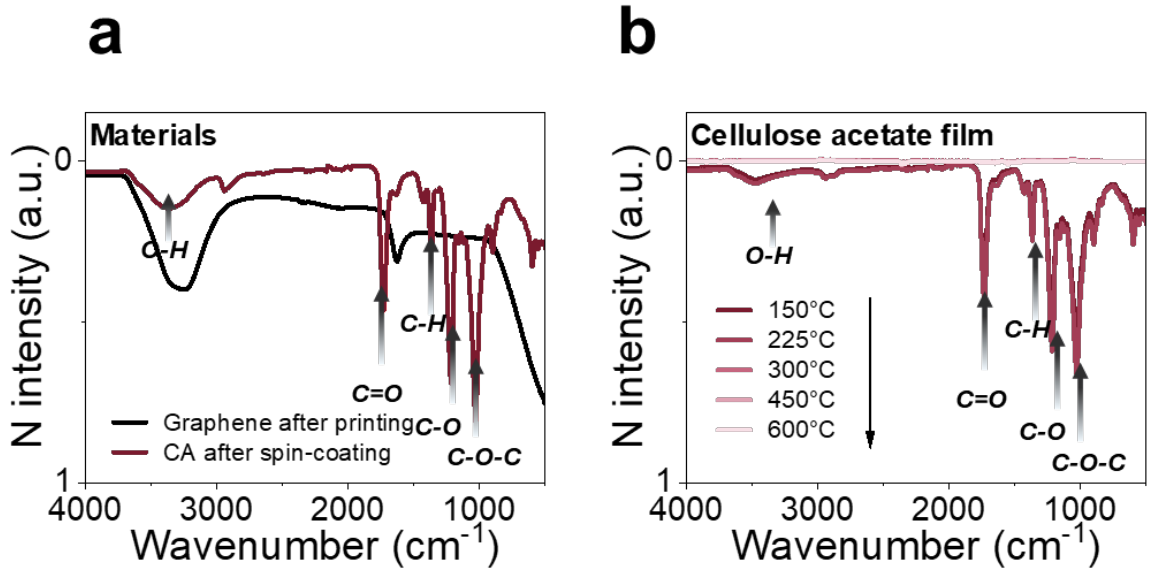

**Figure S16** (a) FTIR scan from raw materials before being degraded. (b) FTIR after degradation at different temperature treatments of the predominant material, cellulose acetate.

**Supplementary Note 1. Mechanism of femtosecond laser ablation in graphene and cellulose acetate.**

Femtosecond laser ablation enables precise patterning of multilayered materials by leveraging distinct energy transfer mechanism unique to each material. In graphene, ablation occurs via non-linear optical absorption, where the ultrafast laser pulses induce rapid multiphoton ionization and localized plasma formation, enabling sharp and energy-efficient removal of conductive materials.[4] In contrast, cellulose acetate undergoes thermally-driven ablation, in which the absorbed laser energy is converted into heat, inducing localized melting and vaporization. The thermal nature of cellulose acetate ablation results in broader lateral material removal due to heat diffusion. [5]

As shown in **Figure 1** of the main text, two fabrication strategies were employed:

(a) **Simultaneous ablation of both graphene and cellulose acetate**, in which both layers are simultaneously patterned by targeting the graphene surface. This patterning method is suitable for horizontal OECTs requiring both channel separation and direct graphene exposure.

(b) **Selective ablation of cellulose acetate**, achieved by tuning the laser parameters to ablate only the top cellulose acetate layer while preserving the underlying graphene. This method is particularly relevant for fabricating vertical OECTs and planar gates.

**Figure S9** highlights the different ablation mechanisms and illustrates how the presence of graphene underneath enhances the ablation efficiency of cellulose acetate. Microscope images show that, the presence of graphene, laser ablation results in heat-related effects like surface bulging. This is attributed to the strong graphene absorption in the visible and near-IR regions and high thermal conductivity, rapidly heating the film during laser exposure. The heat is transferred upward, increasing the temperature of the overlying cellulose acetate, and thereby enlarging the ablation area while inducing heat-related effects like surface bulging. In the absence of graphene, this pronounced ablation effect disappears, and cellulose acetate exhibits only minimal ablation under identical conditions (**Figure S9a**).

The dependence of cellulose acetate's lateral opening profile on this graphene-induced thermal effects further supports the distinction between graphene's non-linear optical-driven ablation and cellulose acetate's thermal-driven ablation. As shown in **Figure S9b,** post-ablation preview of the cellulose acetate surface depicts the heat-induced bulging and expansion effects.

Selective ablation is achieved by carefully tuning both laser fluence and Z-height (focal position) to ensure the energy delivered exceeds the ablation threshold of cellulose acetate while remaining below that of graphene. Cellulose acetate ablation typically initiates at powers of 300–400 mW (depending on film thickness), whereas graphene ablation begins at around 700–800 mW when using a 4× objective. Operating within this power window allows for clean removal of cellulose acetate without damaging the graphene layer. While Z-height adjustment primarily governs spatial precision, laser power remains the dominant parameter that influences the extent of lateral opening in the cellulose acetate layer.

**Supplementary Note 2. 3D Optical Profilometry Data Processing: Algorithm for automated feature recognition and ablation features dimension measurement from surface profiles.**

To quantify ablation outcomes from varying laser ablation parameters, we employed 3D optical profilometry based on coherence scanning interferometry (CSI) to capture Z-height maps of patterned samples. A custom algorithm was developed to extract key geometric features from these scans by analyzing first- and second-order derivatives of the surface profiles.

The surface height profiles obtained from the raw CSI data was processed to extract key morphological features, with feature detection performed via first and second derivative analysis of the Z-height profile. It begins by loading .datx files, decoding their HDF5 structure, and extracting the Z-height data while converting the pixel data into micrometers scale. The pixel size is retrieved from the metadata to ensure accurate spatial scaling.

The colormap data were divided into a central 50% Y-region to be extracted for averaging to focus on the representative region and minimize edge effects. Feature detection was performed by calculating the first and second derivatives of the Z-height profile. The first derivative was used to manually identify local maxima and minima, while the second derivative was employed to detect inflection points to define key surface features.

Separate feature recognition algorithms were applied for cellulose acetate (CA) and graphene features respectively. For the CA features, CA1 (left) was defined as the local minima of the second derivative from the left, and CA2 (right) was the local maxima of the second derivative from the right. For the graphene features, G1 and G2 were identified as local maxima of the Z-height non-derivative data from the left and right sides of the profile, while G0 representing the global minimum of the Z-height data.

The depth of each feature was calculated as the Z-distances from local extrema between the CA1 or CA2 points to G1 or G2 points, and the G1 or G2 points to G0 points. The lateral opening was determined by averaging the X-distance between the CA1 and CA2 points or between the G1 and G2 points across 50% of the respective depth.

**Figure S4a** presents the Z-height colormap, representing the ablated sample's surface profile. The Z-height profile across the entire X-distance at specific Y-levels (with single Y line as shown in **Figure S4b**) was compiled to produce an averaged Z-height profile over the central 50% of the Y-region, illustrated in **Figure S4c**. On this measurement example, the averaged X-distance between G1 and G2 is 13.55 ± 2.39 µm, followed by CA lateral opening of 177.39 ± 12.69 µm and CA depth of 14.44 µm. During the 3D optical profilometry data analysis, microscope images such as illustrated in **Figure S4d** are used alongside the output of automated results to verify valid openings.

**Supplementary References**